\documentclass[traditabstract]{aa} 

\usepackage{graphicx}
\usepackage{txfonts}
\usepackage{longtable}

\usepackage{epsf}
\usepackage{rotating}
\usepackage{graphics}

\def \MJ{M$_{\mathrm{Jup}}$}
\def \RJ{R$_{\mathrm{Jup}}$}

\def \MS{M$_{\odot}$}
\def \kms{km\,s$^{-1}$}
\def \ms{m\,s$^{-1}$}
\def \1s{$1\,\sigma$}
\def \kid{$\chi^2$}
\def \t0{T$_0$}

\newcommand{\teff}{\mbox{$T_{\rm eff}$}}

\newcommand{\vsini}{\mbox{$v \sin i_*$}}
\newcommand{\mictrb}{\mbox{$\xi_{\rm t}$}}
\newcommand{\mactrb}{\mbox{$v_{\rm mac}$}}

\newcommand{\halpha}{\mbox{H$_\alpha$}}

\begin{document}

   \title{WASP-52b, WASP-58b, WASP-59b, and WASP-60b: \\four new transiting close-in giant planets}

   \author{G.~H\'ebrard\inst{1,2}, 
		A.~Collier~Cameron\inst{3},
		D.\,J.\,A.~Brown\inst{3}, 
		R.\,F.\,D\'{\i}az\inst{4,1,2},
		F.~Faedi\inst{13},
		B.~Smalley\inst{5},
   		D.\,R.~Anderson\inst{5}, 
   		D.~Armstrong\inst{13}, 
   		S.\,C.\,C.~Barros\inst{4}, 
		J.~Bento\inst{12}, 
		F.~Bouchy\inst{1,2}, 
		A.\,P.~Doyle\inst{5},
		B.~Enoch\inst{3},
		Y.~G\'omez~Maqueo~Chew\inst{13,14}, 
		\'E.\,M.~H\'ebrard\inst{2}, 
		C.~Hellier\inst{5}, 
		M.~Lendl\inst{6}, 
		T.\,A.~Lister\inst{11}, 
		P.\,F.\,L.~Maxted\inst{5}, 
		J.~McCormac\inst{8}, 
		C.~Moutou\inst{4}, 
		D.~Pollacco\inst{13}, 
		D.~Queloz\inst{6}, 
		A.~Santerne\inst{4,2}, 
		I.~Skillen\inst{8}, 
		J.~Southworth\inst{5},
		J.~Tregloan-Reed\inst{5},
		A.\,H.\,M.\,J.~Triaud\inst{6}, 
		S.~Udry\inst{6}, 
		M.~Vanhuysse\inst{10},
		C.\,A.~Watson\inst{7},
		R.\,G.~West\inst{9},
		P.\,J.~Wheatley\inst{13}}

   \institute{
Institut d'Astrophysique de Paris, UMR7095 CNRS, Universit\'e Pierre \& Marie Curie, 
98bis boulevard Arago, 75014 Paris, France 
\email{hebrard@iap.fr}
\and
Observatoire de Haute-Provence, CNRS/OAMP, 04870 Saint-Michel-l'Observatoire, France
\and
School of Physics and Astronomy, University of St Andrews, St Andrews, Fife KY16 9SS, UK
\and
Laboratoire\,d'Astrophysique\,de\,Marseille,\,Univ.\,de\,Provence,\,CNRS\,(UMR6110),\,38\,r.\,F.\,Joliot\,Curie,\,13388\,Marseille\,cedex\,13,\,France
\and
Astrophysics Group, Keele University, Staffordshire, ST5 5BG, UK
\and
Observatoire de Gen\`eve,  Universit\'e de Gen\`eve, 51 Chemin des Maillettes, 1290 Sauverny, Switzerland
\and
Astrophysics\,Research\,Centre,\,School\,of\,Mathematics\,and\,Physics,\,QueenÕs\,University\,Belfast,\,University\,Road,\,Belfast\,BT7\,1NN,\,UK
\and
Isaac Newton Group of Telescopes, Apartado de Correos 321, 38700 Santa Cruz de Palma, Spain
\and
Department of Physics and Astronomy, University of Leicester, Leicester, LE1 7RH, UK
\and
OverSky, 47, all\'ee des Palanques, BP\,12, 33127 Saint-Jean d'Illac, France
\and
Las Cumbres Observatory, 6740 Cortona Drive Suite 102, Goleta, CA 93117, USA
\and
The Open University, Walton Hall, Milton Keynes, MK7 6AA, UK
\and
Department of Physics, University of Warwick, Gibbet Hill Road, Coventry  CV4 7AL, UK
\and
Department of Physics and Astronomy, Vanderbilt University, 6301 Stevenson Center, 
Nashville, TN 37235, USA
}

   \date{Received TBC; accepted TBC}
      
  \abstract{ We present the discovery of four new transiting hot jupiters, detected mainly from SuperWASP-North 
  and SOPHIE observations. These new planets, WASP-52b, WASP-58b, WASP-59b, and WASP-60b, have 
  orbital periods ranging from 1.7 to 7.9~days, masses between 0.46 and 0.94\,\MJ, and radii between 
  0.73 and 1.49\,\RJ.    
  Their G1 to K5 dwarf host stars have $V$ magnitudes in the range $11.7-13.0$. 
  The depths of the transits are between 0.6 and 2.7\,\%, depending on the target. 
  With their large radii, WASP-52b and 58b are new cases of low-density, inflated planets, whereas 
   WASP-59b is likely to have a large, dense core. WASP-60 shows shallow transits.
   In the case of WASP-52 we also detected the Rossiter-McLaughlin anomaly via time-resolved spectroscopy of a transit. We measured the sky-projected obliquity $\lambda = 24^{\circ}$$^{+17}_{-9}$, 
indicating 
that WASP-52b orbits in the same direction as its host star is rotating and that this prograde 
orbit is slightly misaligned with the stellar equator. 
   These four new planetary systems increase our statistics on hot jupiters, and 
   provide new targets for follow-up studies. }

   \keywords{Planetary systems -- Techniques: radial velocities -- Techniques: photometric --
     Stars: individual: WASP-52, WASP-58, WASP-59, WASP-60}

  \authorrunning{H\'ebrard et al.}
\titlerunning{WASP-52, 58, 59, 60
}

   \maketitle 


\section{Introduction}

About 200 exoplanets are known today as transiting in front of their host stars as
seen from the Earth. That population is particularly interesting since it allows 
numerous studies, including 
accurate radius, mass, and density measurements,
atmospheric studies in absorption through transits and in emission through occultations,
dynamic analyses through possible timing variations,
and obliquity measurements. 
Increasing the size of that sample is essential in order both to improve our statistical 
knowledge on exoplanets, and to discover individual cases particularly well-adapted 
to follow-up studies. Several ground-based photometric programs are surveying 
large fields with that goal, including SuperWASP (Pollacco et al.~\cite{pollacco06})
and HAT (Bakos et al.~\cite{bakos07}). 
More recently, the dedicated space-based missions CoRoT (Baglin et al.~\cite{baglin09}) 
and Kepler (Borucki et al.~\cite{borucki10}) also joined that effort and have discovered 
new transiting exoplanets, in particular with longer periods and smaller radii than are easily
detected by ground-based surveys. Because they survey smaller fields, however,
CoRoT and Kepler detect planets transiting mainly in front of fainter stars, typically 
in the magnitude range $13 < V < 16$. 
Such faint stars make follow-up observations difficult.
By comparison, the planets detected by SuperWASP or HAT transit 
brighter stars, typically in the range $10 < V < 13$, allowing easier and more accurate 
complementary studies. Thus, ground-based surveys for detection of transiting exoplanets remain~pertinent, and are complementary to their space-based~counterparts.

\begin{figure}[ht!] 
\begin{center}
\includegraphics[scale=1.05]{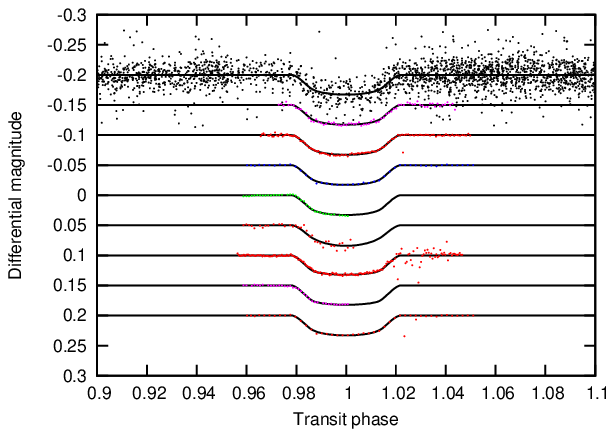}
\includegraphics[scale=1.05]{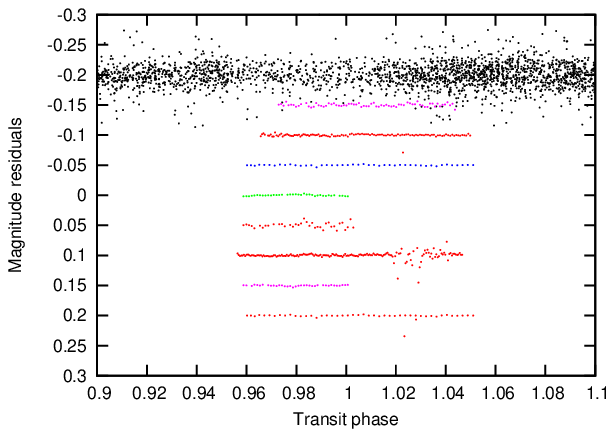}
\includegraphics[scale=0.42]{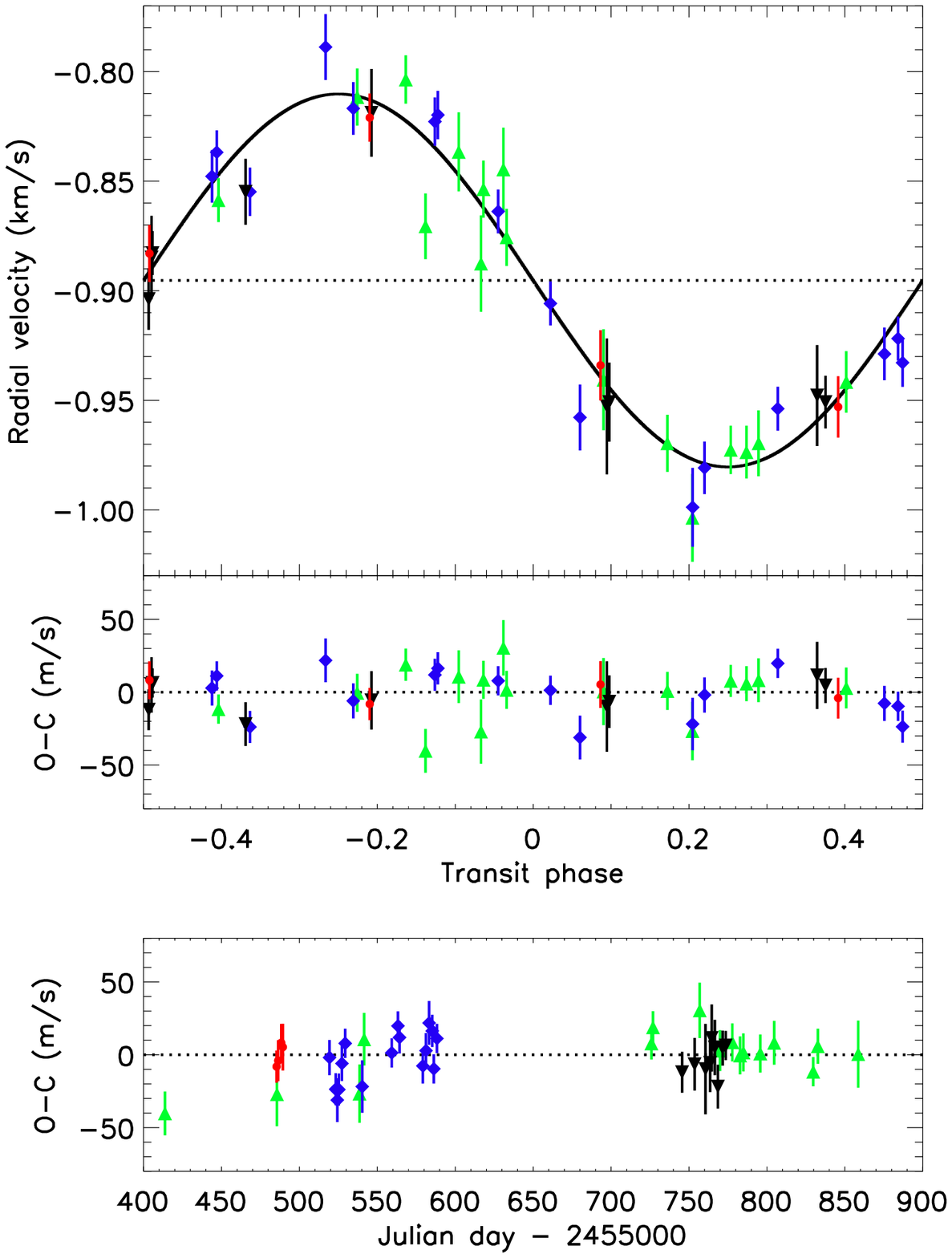}
\caption{Photometry and radial velocities of WASP-52.
The two upper panels show the transit light curves and their residuals 
from the fit.
Different photometric observations are ranked chronologically from top to bottom, 
with the same order as in Table~\ref{table_photom_obs}.
The three lower plots respectively show the phase-folded radial velocities, and their O-C  residuals 
from the Keplerian fit as a function of transit phase and Julian date. 
SOPHIE HE1, HE2 and HR, and CORALIE data (see Table~\ref{table_rv} 
and Sect.~\ref{sect_sophie}) are plotted in red circle, blue diamonds, black downward triangles, and 
green upward triangles, respectively. 
}
\label{fig_obs_W52}
\end{center}
\end{figure}

\begin{figure}[ht!] 
\begin{center}
\includegraphics[scale=1.05]{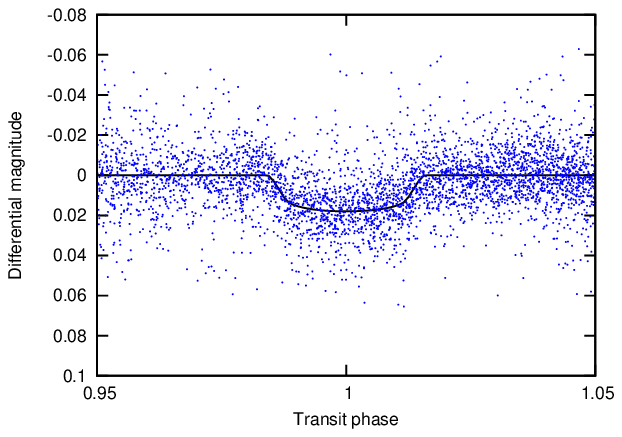}
\includegraphics[scale=1.05]{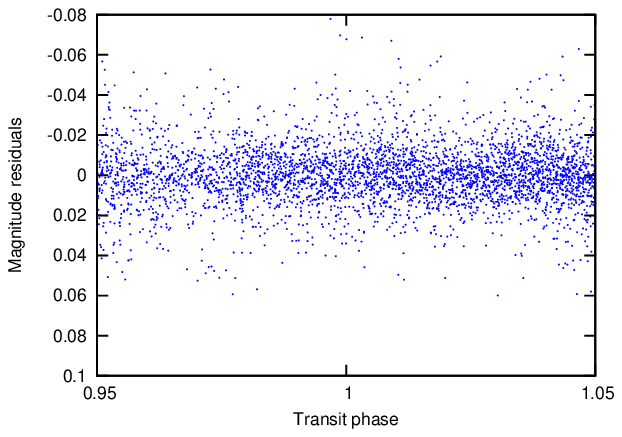}
\includegraphics[scale=0.42]{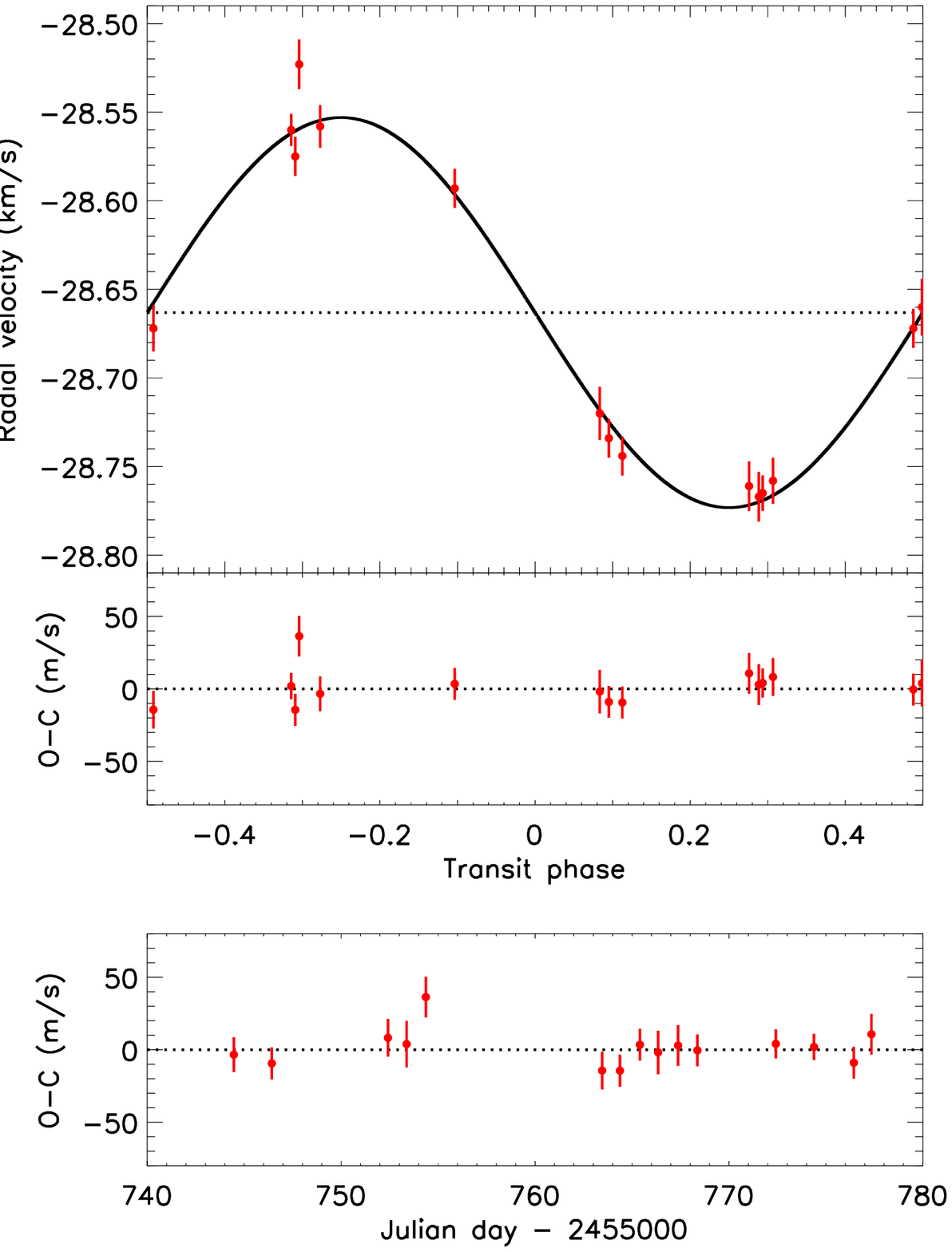}
\caption{Photometry and radial velocities of WASP-58.
The two upper panels show the SuperWASP-N light curve and its residuals 
from the fit. The three lower plots show the SOPHIE radial velocities and their 
residuals from the Keplerian fit. 
}
\label{fig_obs_W58}
\end{center}
\end{figure}

\begin{figure}[ht!] 
\begin{center}
\includegraphics[scale=1.05]{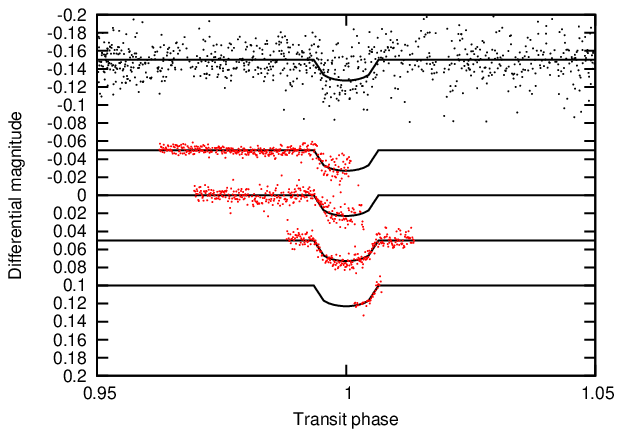}
\includegraphics[scale=1.05]{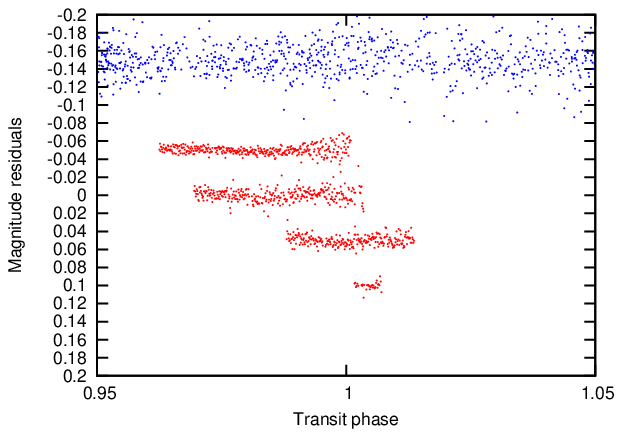}
\includegraphics[scale=0.42]{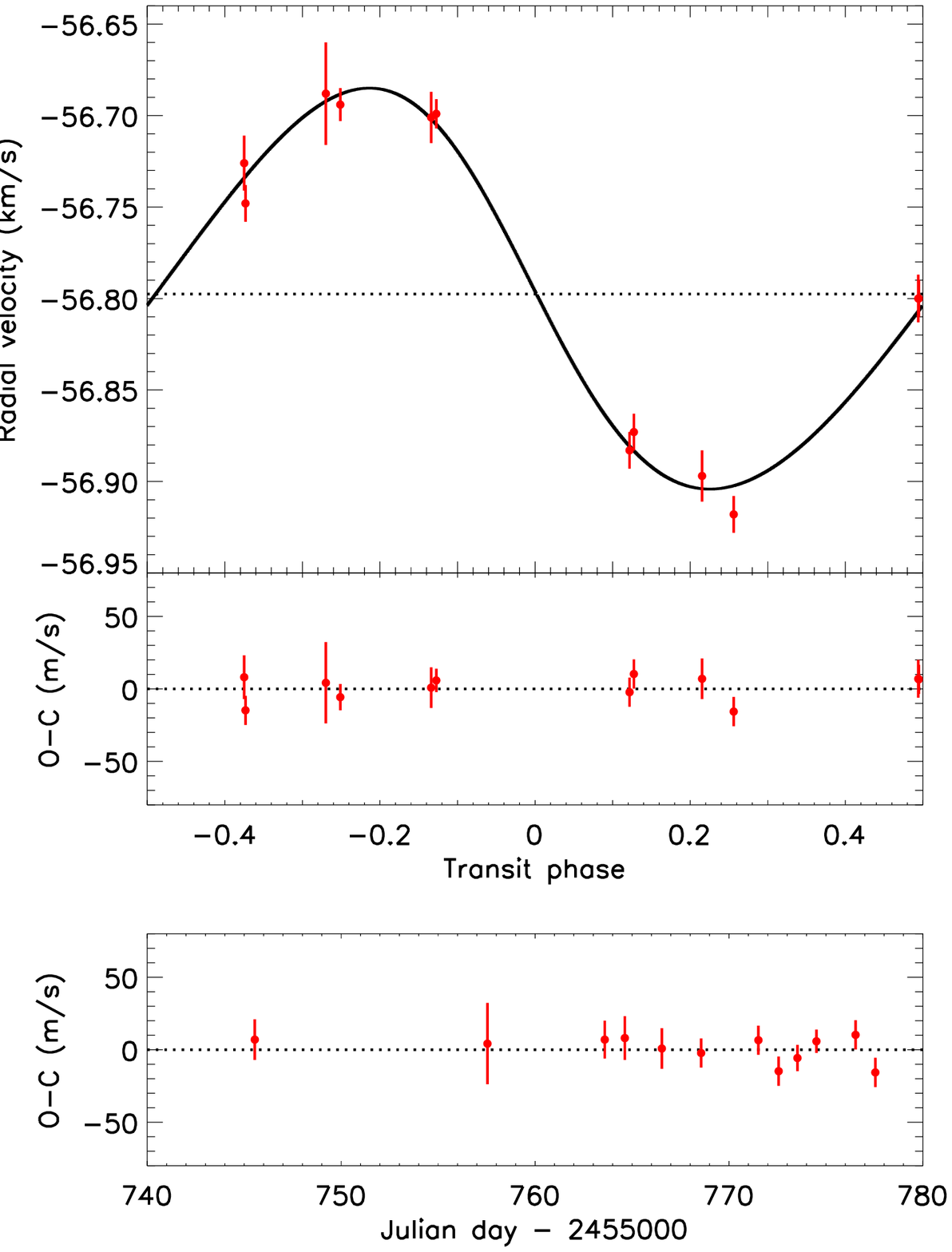}
\caption{Photometry and radial velocities of WASP-59.
The two upper panels show the transit light curves and their residuals 
from the fit.
Different photometric observations are ranked chronologically from top to bottom, 
with the same order as in Table~\ref{table_photom_obs}.
The three lower plots show the SOPHIE radial velocities and their 
residuals from the Keplerian fit. }
\label{fig_obs_W59}
\end{center}
\end{figure}

\begin{figure}[ht!] 
\begin{center}
\includegraphics[scale=1.05]{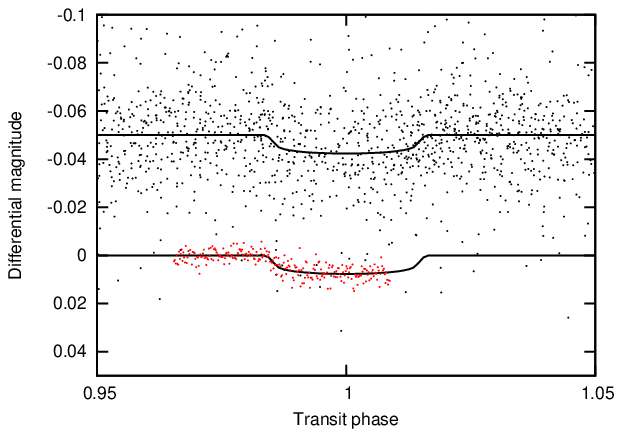}
\includegraphics[scale=1.05]{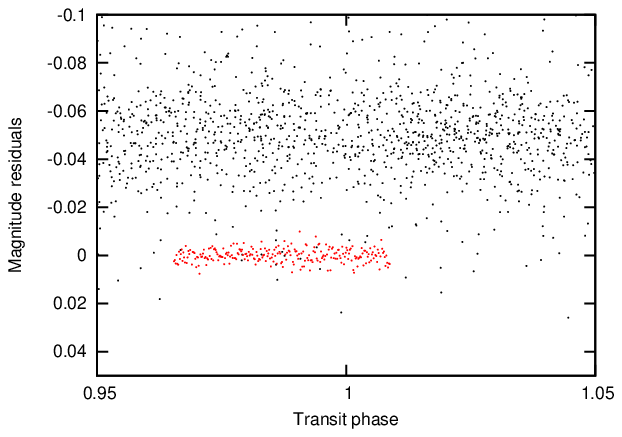}
\includegraphics[scale=0.42]{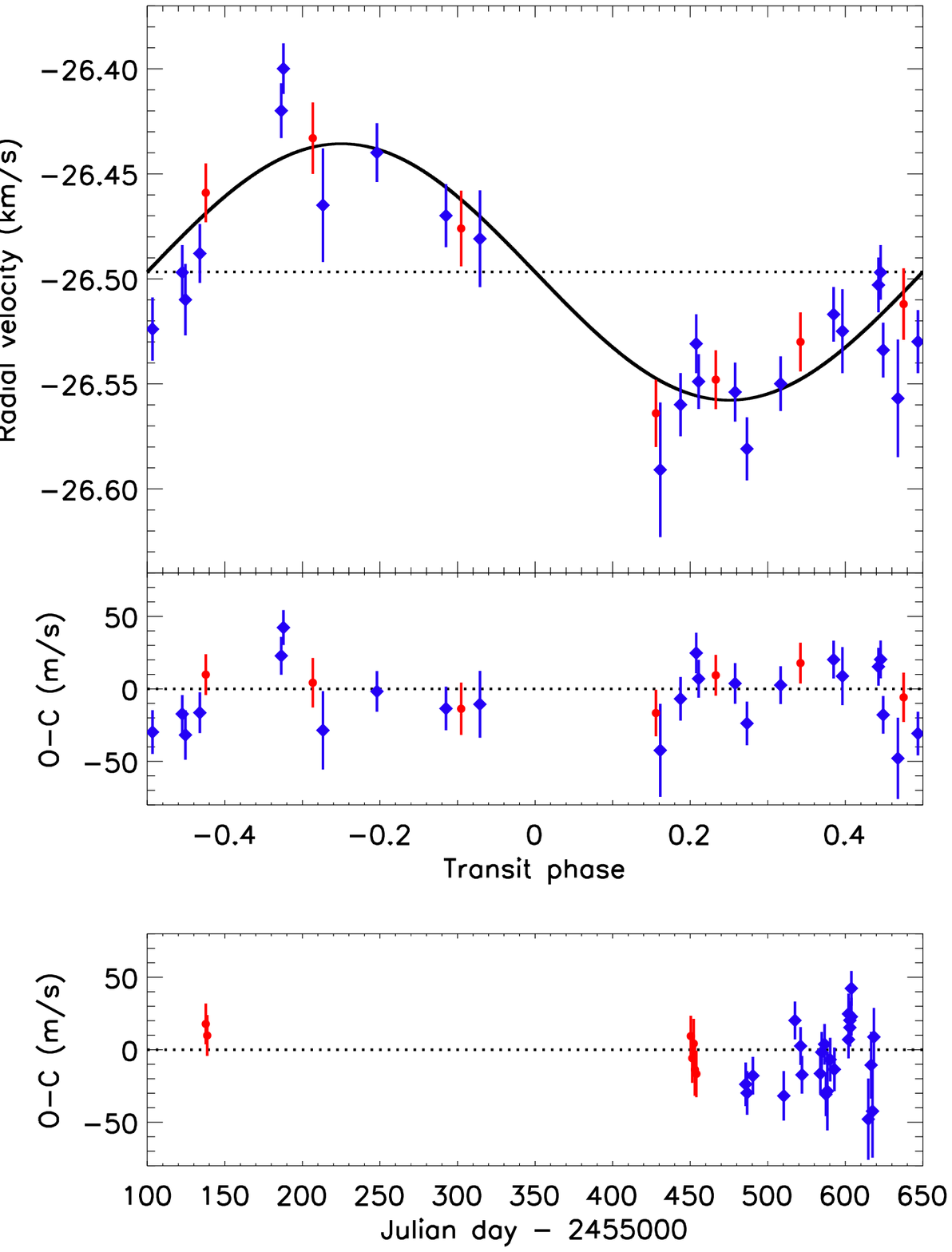}
\caption{Photometry and radial velocities of WASP-60.
The two upper panels show the SuperWASP-N (in black) and NITES (in red) 
transit light curves and their residuals from the fit. 
The three lower plots show the SOPHIE radial velocities and their 
residuals from the Keplerian fit. 
SOPHIE HE1 and H2  (see Table~\ref{table_rv} 
and Sect.~\ref{sect_sophie}) are plotted in red circles and blue diamonds, respectively. 
}
\label{fig_obs_W60}
\end{center}
\end{figure}

We present here the discovery of four new transiting planets, detected with
SuperWASP-North and SOPHIE. These two instruments allowed the detection of the first three 
WASP planets  (Collier Cameron et al.~\cite{cameron07a}; Pollacco et al.~\cite{pollacco08}), 
followed by about twenty others, including inflated hot jupiters such as 
WASP-12b (Hebb et al.~\cite{hebb09}) and WASP-48b (Enoch et al.~\cite{enoch11}), 
compact sub-hot-jupiters like WASP-11b (West et al.~\cite{west09}), 
hot jupiters on eccentric orbits like WASP-38b (Barros et al.~\cite{barros11}), 
and low-mass giant planets, e.g. WASP-13b (Skillen et al.~\cite{skillen09}) 
and WASP-21b (Bouchy et al.~\cite{bouchy10}).

The four new planets presented here
are 
\object{WASP-52b},
\object{WASP-58b},
\object{WASP-59b}, and
\object{WASP-60b}.
They are giant, close-in planets, with 
masses ranging from  0.46 to 0.94\,\MJ, 
orbital periods between 1.7 and 7.9~days, 
and transit depths between 0.6 and 2.7\,\%.
We describe in Sect.~\ref{sect_observations} the observations that allowed their discovery, 
their analyses in Sect.~\ref{sect_analysis}, and conclude in~Sect.~\ref{sect_concle}.

\section{Observations}
\label{sect_observations}

\subsection{Photometric detection with SuperWASP}

The coordinates and magnitudes of the stars  WASP-52, 58, 59, and 60 are reported 
in the upper part of Table~\ref{table_stell_params}. They were first identified as promising 
candidates host stars for transiting planets with SuperWASP-North. Located on La Palma
in the Canary Islands, Spain, it consists of eight Canon 200\,mm $f/1.8$ focal lenses coupled to 
e2v $2048 \times 2048$~pixel CCDs with 13.7" pixels and a field-of-view of 
$7.8^{\circ} \times 7.8^{\circ}$, associated with a custom-built photometric reduction pipeline
(Pollacco et al.~\cite{pollacco06}). It observes with a broad-band custom filter
(400-700~nm).
Thousands of photometric points were secured with 
SuperWASP-North over several seasons for each target (see 
Table~\ref{table_photom_obs} for a~summary).

\begin{table*}[ht!]
\begin{center}
\caption{Stellar parameters of the four planet host stars from analysis of SOPHIE spectra.}
\begin{tabular}{lrrrr} \hline
Parameter  		&   WASP-52 		&   WASP-58		&  WASP-59 		&	WASP-60				\\ 
\hline
RA (J2000)		&      23:13:58.76	&       18:18:48.25	&    23:18:29.54	&	23:46:39.98			\\
DEC (J2000)		&   $+$08:45:40.6	&    $+$45:10:19.1	& $+$24:53:21.4		&    $+$31:09:21.4			 \\	
$B$			&	12.9		&	12.33		&	13.92		&	12.86				\\
$V$			&     12.0		&	11.66		&	13.00		&	12.18				\\
\hline
\teff\ (K)     		&   $5000 \pm 100$ 	&   $5800 \pm 150 $	&  $4650 \pm 150$ 	&	$5900 \pm 100$			\\
$\log g_*$      		&   $4.5 \pm 0.1$ 	&   $4.3 \pm 0.1 $	&  $4.55 \pm 0.15$    	&	$4.2 \pm 0.1$			\\
\mictrb\ (\kms)    	&   $0.9 \pm 0.1$ 	&   $1.1 \pm 0.2 $	&  $0.7 \pm 0.3$ 	&	$1.0 \pm 0.1$			\\
\mactrb\ (\kms) 	&   $0.7 \pm 0.3$ 	&   $2.3 \pm 0.3$	&  $0	 $ 	  	&	$2.9 \pm 0.3$			\\
\vsini\ (\kms)     	&   $3.6 \pm 0.9$ 	&   $2.8 \pm 0.9 $	&  $2.3 \pm 1.2$ 	&	$3.4 \pm 0.8$			\\
$P_{\rm rot}$ (d)	&   $11.8 \pm 3.3$	&   $21.1 \pm 7.6$ 	&  $15.9 \pm 8.4$	&	$20.6 \pm 5.5$			\\
Gyrochron. age (Gyr)		&   $0.4^{+0.3}_{-0.2}$	&   $3.2^{+4.5}_{-2.1}$ &  $0.5^{+0.7}_{-0.4}$	&	$3.6^{+4.3}_{-2.1}$		\\
$\log{R'_\mathrm{HK}}$  &   $ -4.4 \pm 0.2$ 	&   $ -4.4 \pm 0.2 $	&  $-4.1 \pm 0.2$	& $ -4.4 \pm 0.2 $			\\
{[Fe/H]}   		&   $0.03 \pm 0.12$ 	&   $-0.45 \pm 0.09$	&  $-0.15 \pm 0.11$ 	& 	$-0.04 \pm 0.09$		\\
{[Na/H]}   		&   $0.29 \pm 0.08$ 	&   $-0.44 \pm 0.07$ 	&	--		&	$0.07 \pm 0.11$			\\
{[Mg/H]}   		&   $0.26 \pm 0.13$ 	&   $-0.30 \pm 0.07$ 	&	--		&	$0.02 \pm 0.06$			\\
{[Si/H]}   		&   $0.25 \pm 0.18$ 	&   $-0.34 \pm 0.06$ 	&	--		&	$0.15 \pm 0.08$			\\
{[Ca/H]}   		&   $0.04 \pm 0.13$ 	&   $-0.30 \pm 0.13$ 	&  $0.23 \pm 0.18$ 	&	$0.06 \pm 0.13$			\\
{[Sc/H]}   		&   $0.17 \pm 0.11$ 	&   $-0.22 \pm 0.07$ 	&	--		&	$0.18 \pm 0.08$			\\
{[Ti/H]}   		&   $0.06 \pm 0.12$ 	&   $-0.29 \pm 0.07$ 	&  $0.28 \pm 0.18$ 	&	$0.00 \pm 0.07$			\\
{[V/H]}    		&   $0.19 \pm 0.13$ 	&	--		&  $0.66 \pm 0.19$ 	&	$0.02 \pm 0.10$			\\
{[Cr/H]}   		&   $0.11 \pm 0.15$ 	&   $-0.39 \pm 0.18$ 	&	--		&	$0.05 \pm 0.14$			\\
{[Co/H]}   		&   $0.20 \pm 0.07$ 	&	--		&  $0.16 \pm 0.11$ 	&	$-0.01 \pm 0.15$		\\
{[Ni/H]}   		&   $0.11 \pm 0.13$ 	&   $-0.49 \pm 0.09$ 	&  $-0.10 \pm 0.10$ 	&	$0.29 \pm 0.08$			\\
log A(Li)  		&   $<0.3\pm0.2 $	&   $2.32 \pm 0.13$ 	&  $<0.17 \pm 0.21$ 	&	$< 0.83 \pm 0.09$		\\
Sp. Type 		&   K2V 		&   G2V 		&  K5V 		  	&	G1V				\\
Distance (pc) 		&   $140 \pm 20$ 	&   $300 \pm 50$ pc 	&  $125 \pm 25$ 	&	$400 \pm 60	$		\\ 
\hline	
\\
\end{tabular}
\label{table_stell_params}
\end{center}
\end{table*}

Periodic signatures of possible planetary transits were identified in these light curves 
using the algorithms presented by Collier Cameron et al.~(\cite{cameron06}). Phase-folded 
SuperWASP light curves are plotted in the upper panels of Figs.~\ref{fig_obs_W52}, 
\ref{fig_obs_W58}, \ref{fig_obs_W59}, and \ref{fig_obs_W60}, for WASP-52, 58, 59, and 
60, respectively. They show dips  of the order of one percent, compatible with 
transiting  giant planets with periods of 1.7, 5.0, 7.9, and 4.3~days, respectively.

\subsection{Radial velocities with the spectrograph SOPHIE}
\label{sect_sophie}

The spectroscopic follow-up of these four candidates was done with 
SOPHIE, the spectrograph dedicated to high-precision radial velocity 
measurements at the 1.93-m telescope of Haute-Provence Observatory, 
France (Bouchy et al.~\cite{bouchy09}). 
The first goal of these observations is to establish the planetary nature of the transiting 
candidates found in photometry, which in most cases are not due to planets 
but to astrophysical false positives such as blended stellar binaries.
The second goal is to characterize the secured planets 
by measuring in particular their masses and orbital eccentricities.

SOPHIE was mainly used in High-Efficiency mode (HE) with a resolving power 
$R=40\,000$ in order to increase the throughput for these faint targets.
The exposure times ranged from 10 to 30\,min depending on the targets, and they 
were adjusted as a function of the weather conditions in order to keep the 
signal-to-noise ratio as constant as possible for any given target.
The spectra were extracted using the SOPHIE pipeline, and the radial velocities 
were measured from weighted cross-correlation with numerical masks
characteristic of the spectral type of the target (Baranne et al.~\cite{baranne96}; 
Pepe et al.~\cite{pepe02}). 
We adjusted the number of spectral orders used in the cross-correlation in order 
to reduce the dispersion of the measurements. Indeed, some spectral domains  
are noisy (especially in the blue part of the spectra), so using them degrades the 
accuracy of the radial-velocity~measurement. 

The error bars on the radial velocities were computed from the cross-correlation 
function using the method presented by Boisse et al.~(\cite{boisse10}). 
Some spectra were contaminated by moonlight. 
Following the method described in Pollacco et al.~(\cite{pollacco08}) and
H\'ebrard et al.~(\cite{hebrard08}), we estimated and corrected the moonlight contamination 
by using the second SOPHIE fibre aperture, which is targeted on the sky whereas the first 
aperture points toward the target. This typically results in 
radial velocity corrections of a few~\ms\ 
up to a few tens of~\ms. Removing these points does not significantly modify the orbital solutions.

\begin{figure*}[t!] 
\begin{center}
\includegraphics[scale=0.61]{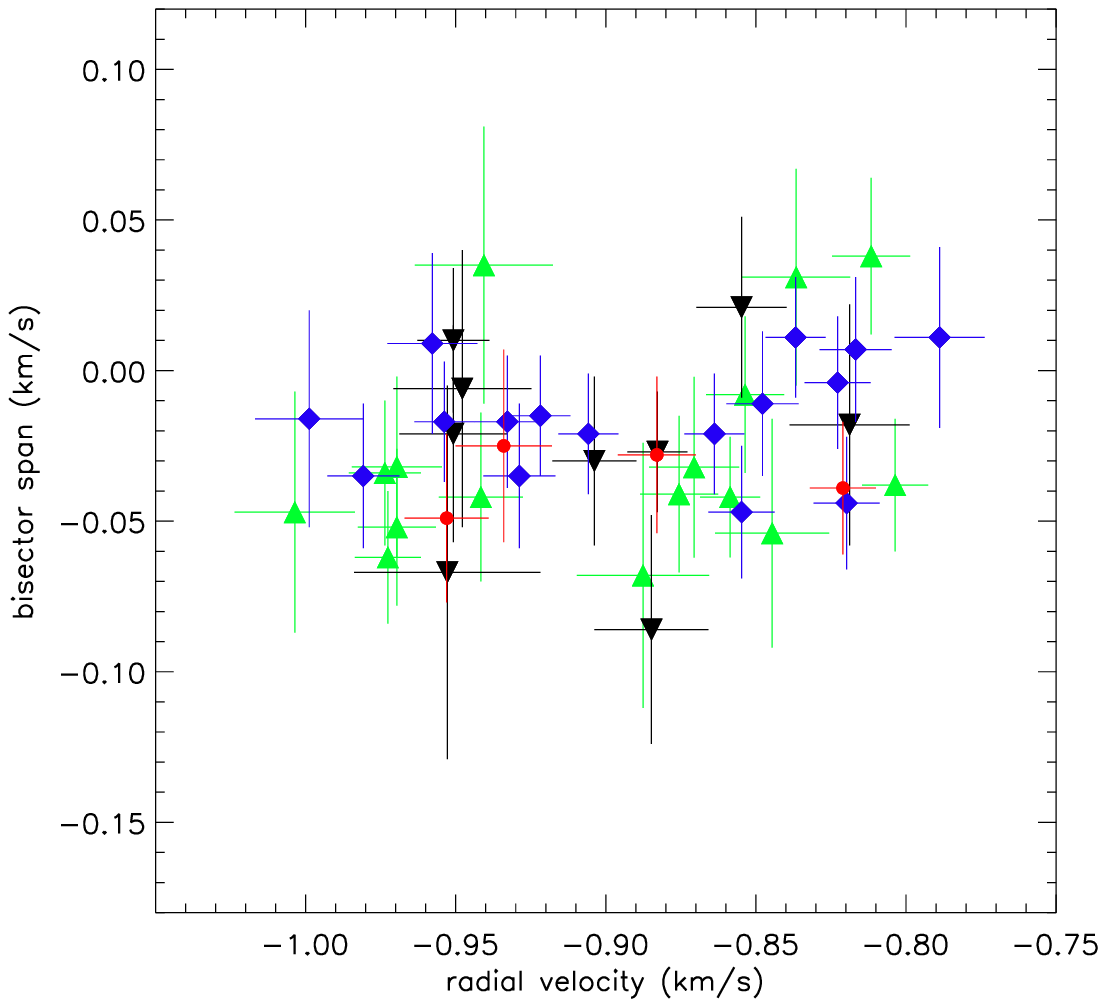}
\includegraphics[scale=0.61]{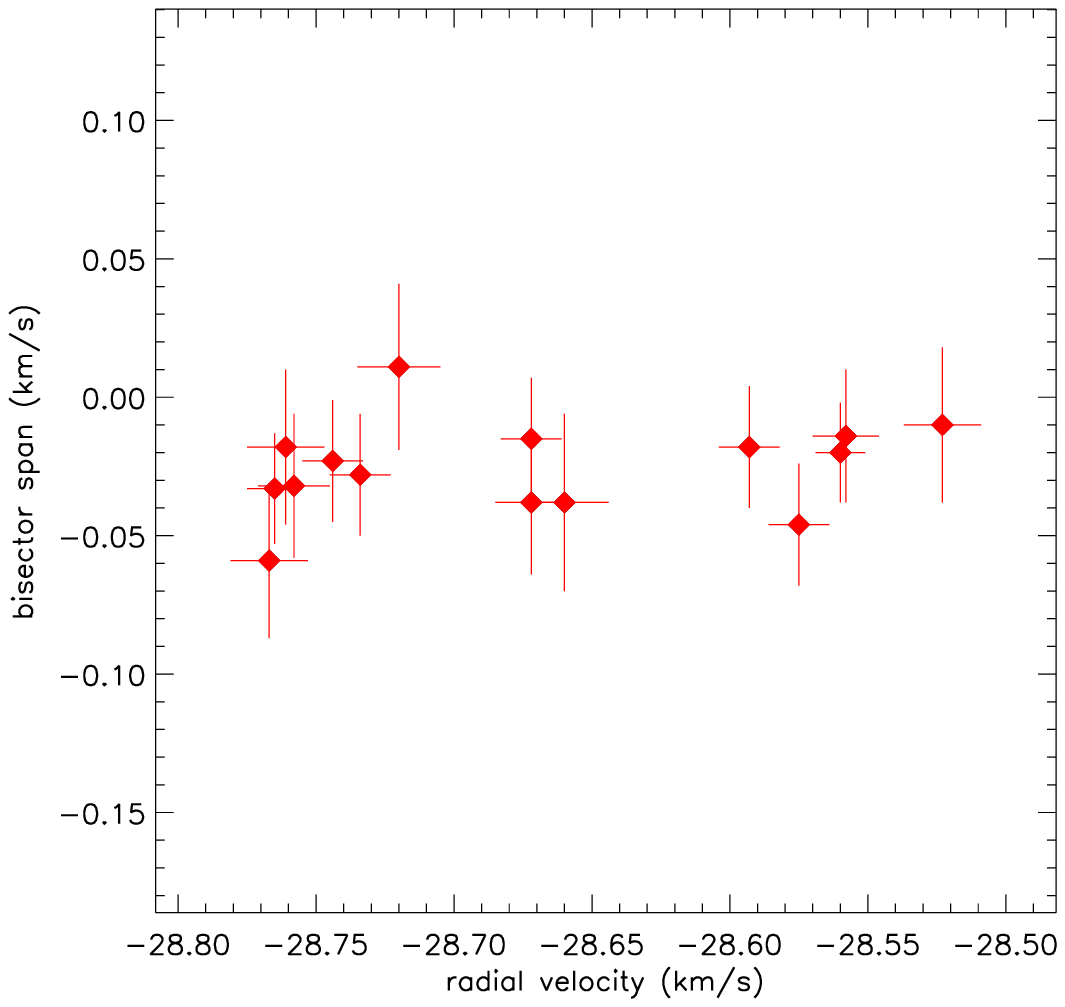}
\includegraphics[scale=0.61]{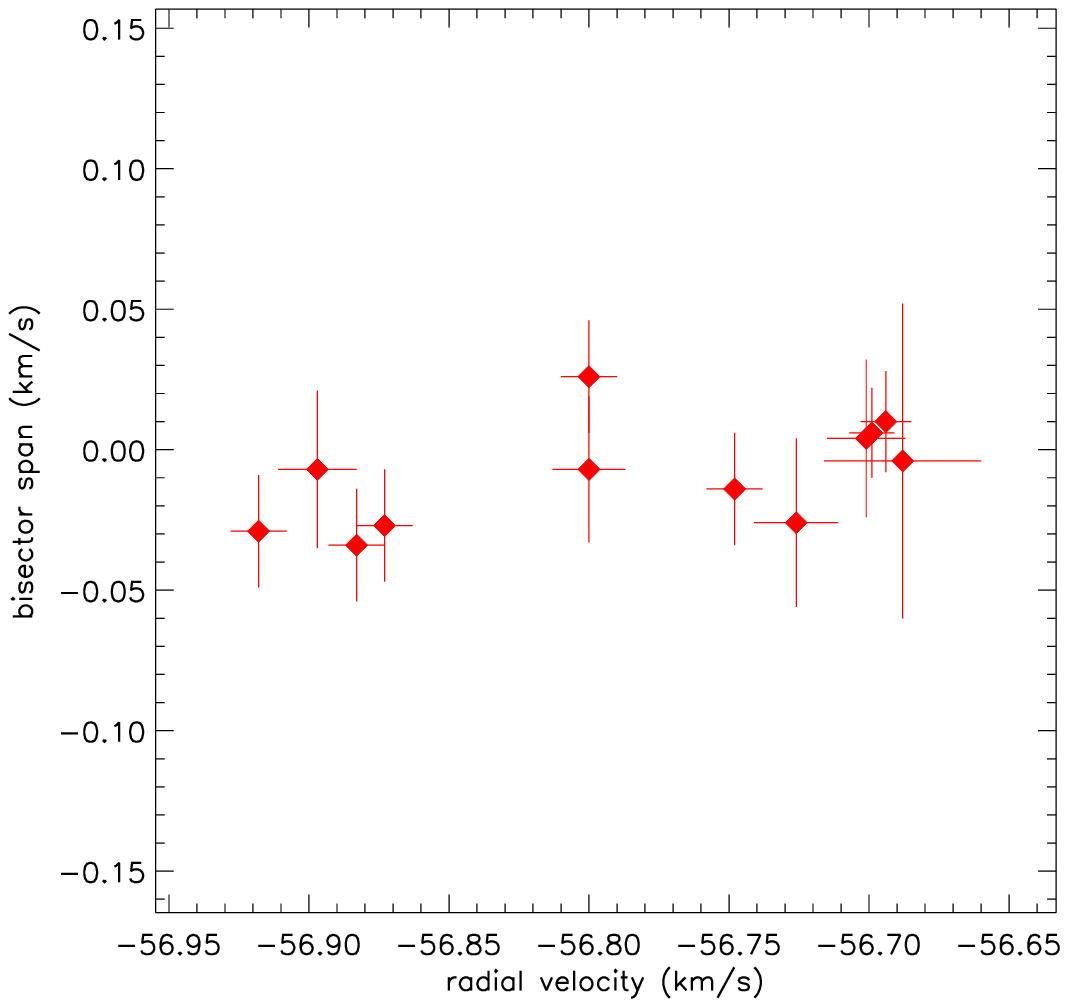}
\includegraphics[scale=0.61]{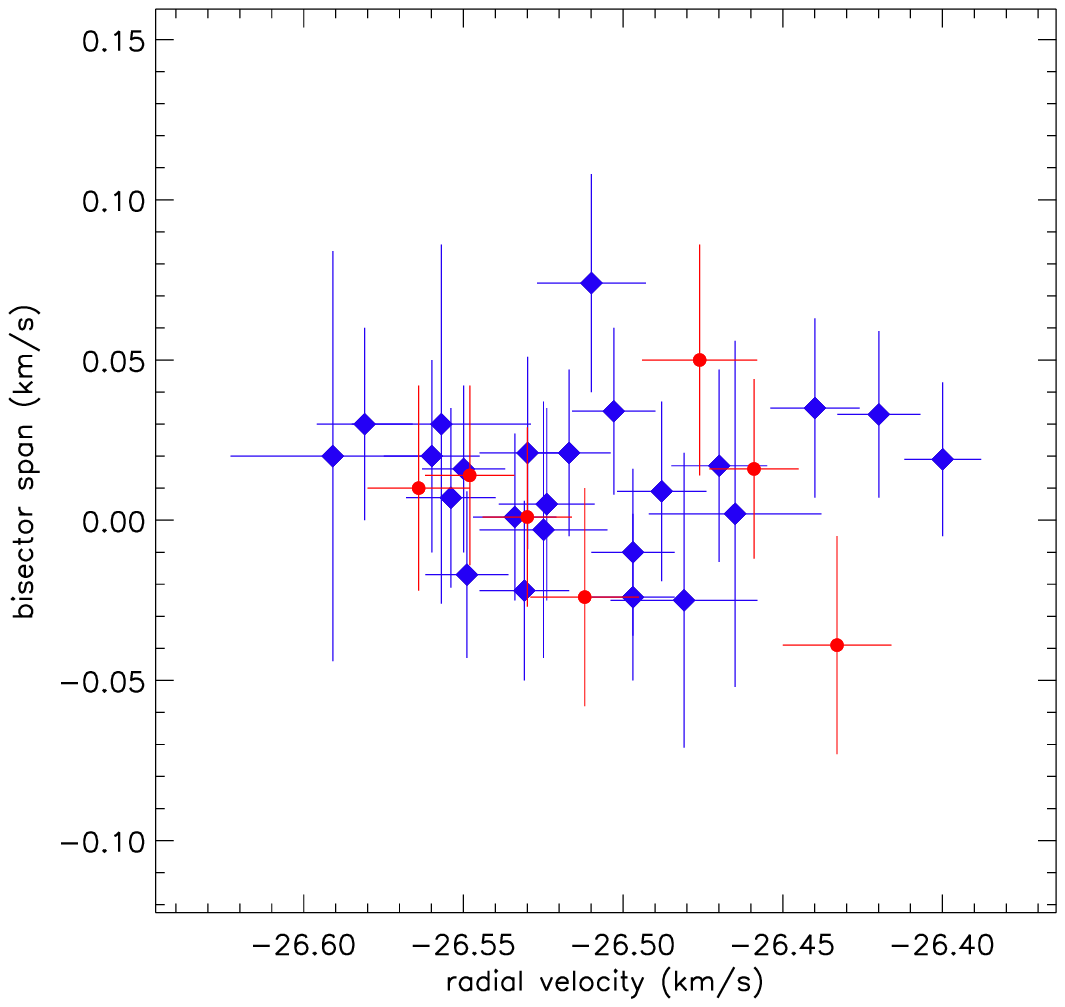}
\caption{Bisector span as a function of the radial velocities with 1-$\sigma$\,Êerror bars
for WASP-52, 58, 59 and 60 (from left to right and top to bottom). Color codes are the 
same as in Figs.~\ref{fig_obs_W52}-\ref{fig_obs_W60}.
The ranges here have the same extents in the $x$- and $y$-axes.
}
\label{fig_bisectors}
\end{center}
\end{figure*}

The radial velocity measurements are reported in Table~\ref{table_rv} and are displayed
in the lower panels of  Figs.~\ref{fig_obs_W52}--\ref{fig_obs_W60}, together with their 
Keplerian fits and the residuals. They show variations in phase with the SuperWASP
transit ephemeris and with semi-amplitudes of a few tens~\ms, implying companion masses 
slightly below a Jupiter mass.
Radial velocities measured using different stellar masks (F0, G2, or K5) produce 
variations with similar amplitudes, so there is no 
evidence that their variations could be explained by blend scenarios 
caused by stars of different spectral types. Similarly, the cross-correlation function bisector
spans show neither variations or trends as a function of radial velocity (Fig.~\ref{fig_bisectors}).
This reinforces the conclusion that the radial-velocity variations are not caused by 
spectral-line profile changes attributable to~blends or stellar activity.
We thus conclude that our four targets harbour transiting giant planets, which we designate as 
WASP-52b, WASP-58b, WASP-59b, and WASP-60b hereafter.

The raw residuals of the Keplerian fits to two of the four targets (WASP-52 and 60) 
show an abrupt offset of $\sim100\,$\ms\ around 
BJD\,$=2\,455\,500$ which could be due to possible extra companions in the planetary 
systems. However, the fact that the offset is similar in amplitude and time 
for both targets suggests it 
could be instrumental. WASP-58 and 59 were not observed at that epoch.
To constrain possible extra companions we re-observed 
WASP-52 the following season with SOPHIE in High-Resolution mode (HR), with a 
resolving power $R=75\,000$ following the installation of new octagonal fiber
sections in the spectrograph, which improved significantly the radial-velocity accuracy of SOPHIE  
 (Bouchy et al.~\cite{bouchy12}). WASP-52 was also 
observed with the CORALIE spectrograph mounted on the 1.2-m Euler-Swiss 
telescope at La Silla (Queloz et al.~\cite{queloz00}).
These new measurements did not confirm any structure in the residuals of the 
one-Keplerian fit, so we conclude that the offsets detected during the previous season on
WASP-52 and 60 were instrumental. 
Such systematic effects have already been observed in the past with the HE mode 
of SOPHIE, but not with the HR mode. They are not well understood, but seem to 
be linked with temperature variations in the telescope dome. To circumvent that 
effect, we considered the SOPHIE HE data of WASP-52 and 60 secured before and 
after the zero-point change as independent datasets, allowing a free radial-velocity offset to be 
fitted between them. These are noted "SOPHIE HE1" and "SOPHIE HE2" in 
Table~\ref{table_rv}.

\subsection{Additional photometry of the transits}

Once the radial velocities established that the four candidates identified with photometry 
actually were transiting planets, we obtained extra transit light curves with telescopes
larger than SuperWASP, allowing improved resolution and more precise time-series 
photometry during transit. 
The goal is to refine the determination of the parameters derived 
from photometry. It was not possible to observe extra transits of WASP-58b with our 
available telescopes, due to the orbital period of that system being almost exactly an integer number   
of days ($P=5.018$~days). 

We used the six telescopes below equipped  with cameras
to observe extra transit light curves of the three remaining planets:

\noindent
\textit{-- JGT:} the 0.94-m James Gregory Telescope located at University of St. Andrews, Scotland;

\noindent
\textit{-- FTS:} the 2-m Faulkes Telescope South  located at Siding Spring Observatory in New South Wales, Australia;

\noindent
\textit{-- EulerCam:} the new camera at the 1.2-m Euler-Swiss telescope at La Silla, Chile;

\noindent
\textit{-- BUSCA:} at the Calar Alto Observatory 2.2-m telescope, Spain;

\noindent
\textit{-- OverSky:} the 0.36-m telescope at the OverSky Observatory, La Palma, Spain;

\noindent
\textit{-- NITES:} a 0.4-m telescope at La Palma, Spain.

The dates of the transit observations made with these facilities as well as the 
filters used  are reported in Table~\ref{table_photom_obs}. 
Standard aperture photometry was carried out on all these datasets, using 
comparison stars available in the different fields for relative photometry.
The extra light curves of WASP-52, 59, and 60 are plotted in the upper panels of 
Figs.~\ref{fig_obs_W52}, \ref{fig_obs_W59}, and \ref{fig_obs_W60}, respectively.

\subsection{Transit spectroscopy}
\label{sect_obs_RM}

A transit of WASP-52b was observed in spectroscopy with SOPHIE.
The goal was to detect the Rossiter-McLaughlin anomaly, which is an 
apparent time-dependent distortion of the stellar line profiles that occurs when a 
planet transits the disc of a rotating star. It allows the measurement of the 
sky-projected angle 
between the planetary orbital axis and the stellar rotation axis, usually denoted 
$\lambda$ (see, e.g., H\'ebrard et al.~\cite{hebrard08}). 
The spectroscopic transit was observed  in the HE mode  of SOPHIE
to improve the~throughput. 

The target was observed continuously during a four-hour sequence under 
good weather conditions on 2011 August 21. 
19 measurements were secured with exposure times 
ranging from 600 to 1200~seconds, including nine within the transit.
The remaining observations obtained before and after the transit are 
mandatory for reference points. 
The radial velocities were extracted as above 
(Sect.~\ref{sect_sophie}), but using slightly fewer orders for 
the cross-correlation to reduce the dispersion of the measurements.
This depends on the flux color balance in the spectra, which depends 
on the airmasses of the different observations.

These observations are plotted in Fig.~\ref{fig_RM_W52} and are labelled
"SOPHIE HE-RM" in Table~\ref{table_rv}.
The Rossiter-McLaughlin anomaly is detected, with a full amplitude of 
$\sim80$\,\ms. The shape of the anomaly indicates a prograde 
orbit, with an asymmetry that suggests misalignment (see Sect.~\ref{sect_fit_RM}).

For different reasons, including poor weather conditions, we were unable to 
perform time-resolved spectroscopy of the transits of the three other planets.

\begin{table}[h]
\caption{Photometric observations.}
\centering
\begin{centering}
\begin{tabular}{lcr} 
\hline
Telescope & Band & Date \\ 
\hline
\multicolumn{3}{l}{\textbf{\hspace{0.7cm}WASP-52}}  \\
SuperWASP-N  	& -	&	2008 Jul - 2009 Nov \\
JGT 		& r  	&	2010 Oct 24		\\ 
FTS 		& z	&	2011 Aug	02		\\ 
EulerCam	& r  	&	2011 Aug 20		\\ 
BUSCA 	& u  	&	2011 Aug 27		\\ 
BUSCA 	& g  	&	2011 Aug 27		\\ 
BUSCA 	& r  	&	2011 Aug 27		\\ 
BUSCA 	& z  	&	2011 Aug 27		\\ 
EulerCam	& r  	&	2011 Sep 24		\\ 
\hline	
\multicolumn{3}{l}{\textbf{\hspace{0.7cm}WASP-58}}  \\
SuperWASP-N  	& -	&	2004 May - 2010 Aug \\ 
\hline	
\multicolumn{3}{l}{\textbf{\hspace{0.7cm}WASP-59}}  \\
SuperWASP-N  	& -	&	2004 May - 2006 Nov \\
OverSky	& r  	&	2011 Oct 28		\\ 
OverSky	& r  	&	2011 Nov 04		\\ 
OverSky	& r  	&	2011 Nov 20		\\ 
JGT 	& r  	&	2011 Nov 20		\\ 
\hline	
\multicolumn{3}{l}{\textbf{\hspace{0.7cm}WASP-60}}  \\
SuperWASP-N  	& -	&	2004 May - 2007 Dec \\
NITES	& r  	&	2011 Aug 25		\\ 
\hline	
\end{tabular}
\end{centering}
\label{table_photom_obs}
\end{table}

\section{Analysis}
\label{sect_analysis}

\subsection{Parameters of the host stars}

Spectral analysis of the four host stars was performed with the co-added 
individual spectra obtained with SOPHIE for radial velocity measurements 
(Sect.~\ref{sect_sophie}). We used here only the spectra without moonlight 
contamination, and only the HE spectra out of transit in the case of WASP-52.
This allows high-enough signal-to-noise ratio to be reached, with a given 
spectral resolution, and avoid any possible contamination in the spectra.
The parameters obtained from these analyses are listed in the bottom part 
of Table~\ref{table_stell_params}.

We used the methods given by
\cite{2009A&A...496..259G}. The \halpha\ line was used to determine the
effective temperature \teff, except for WASP-59 where \halpha\ was too 
weak; for that star \teff\ was measured 
from the null dependence of abundance on excitation potential. 
The surface gravity $\log g_*$ was determined from the Ca\,{\sc i} lines at 
6162\,{\AA} and 6439\,{\AA} (\cite{2010A&A...519A..51B}), 
along with the Na\,{\sc i}~D and Mg\,{\sc i}~b lines.
The spectral types were established from \teff\
using the table in Gray~(\cite{gray08}). 

The elemental abundances
were determined from equivalent width measurements of several clean and
unblended lines. The microturbulence \mictrb\ was evaluated from
Fe\,{\sc i} using the method of \cite{1984A&A...134..189M}. The quoted error
estimates include that given by the uncertainties in \teff, $\log g_*$ and \mictrb,
as well as the scatter due to measurement and atomic data uncertainties.

The projected stellar rotation velocities \vsini\ were determined by fitting the
profiles of several unblended Fe\,{\sc i} lines. They agree with the measurements 
secured from the cross-correlation function following Boisse et al.~(\cite{boisse10}).
The values for macroturbulence \mactrb\ were assumed based on the tabulations 
by Gray~(\cite{gray08}) and \cite{2010MNRAS.405.1907B}.
Macroturbulence was assumed to be zero for WASP-59, since for mid-K stars it is 
expected to be lower than that of thermal broadening (Gray~\cite{gray08}). 
An instrumental FWHM of 0.15 $\pm$ 0.01\,{\AA} was measured from the telluric 
lines around 6300\,\AA, and best fitting values of the \vsini\ were obtained.

Lithium is detected only in WASP-58. The temperature of 5800\,K along with the 
derived lithium abundance $\log$\,A(Li) imply a lower age limit of $\sim2$\,Gyr 
when compared with NGC\,752 (\cite{2005A&A...442..615S}).
There is no significant detection of lithium in the spectra of WASP-52, 59 
and 60, with equivalent width upper limits of 4\,m\AA, 15\,m\AA, and 6\,m\AA, 
respectively. The lack of lithium would suggest lower age limits of 0.5~Gy and several Gyr 
for WASP-52 and 60, respectively (\cite{2005A&A...442..615S}). 
For the mid-K star WASP-59, lithium is expected to be depleted in only a few 100 Myr.

For WASP-52, 58, 59, and 60, the rotation rates implied by the \vsini\ 
($P_{\rm rot} = 11.8 \pm 3.3$~d, 
$21.1 \pm 7.6$~d, 
$15.9 \pm 8.4$~d, and
$20.6 \pm 5.5$~d, respectively)
gives gyrochronological ages of 
$ 0.4^{+0.3}_{-0.2}$~Gyr, 
$ 3.2^{+4.5}_{-2.1}$~Gyr, 
$ 0.5^{+0.7}_{-0.4}$~Gyr, and
$ 3.6^{+4.3}_{-2.1}$~Gyr, respectively, 
using the \cite{2007ApJ...669.1167B} relation. 

WASP-52 and 59 show chromospheric emission peaks in the Ca\,{\sc ii} H+K lines
that indicate they are active stars, whereas WASP-58 and 60 do not 
show any evident emission. We computed the $\log{R'_\mathrm{HK}}$ indices 
following the calibration of Boisse et al.~(\cite{boisse10})

\begin{figure*}[bt!] 
\begin{center}
\includegraphics[width=0.9\textwidth]{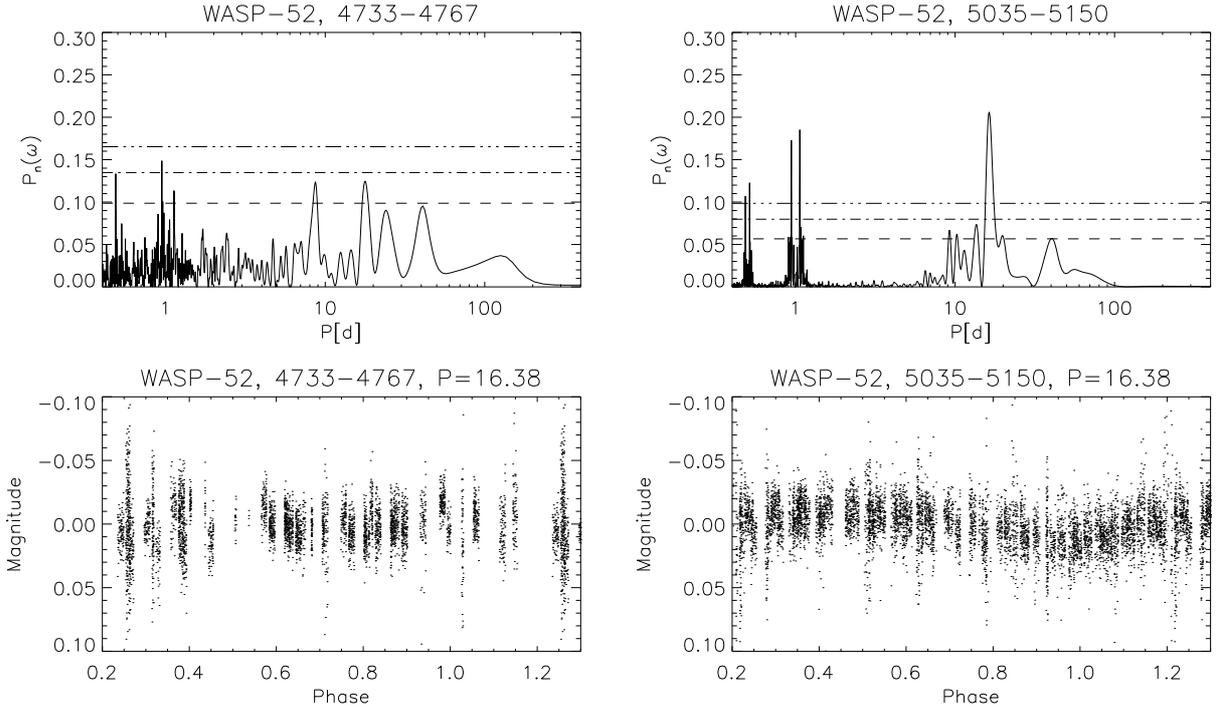}
\end{center}
\caption{{\it Upper panel:} Periodograms for the WASP data from two seasons for
 WASP-52. The date range (JD$-$245000) is given in the title of each panel.
Horizontal lines indicate false alarm probability levels FAP\,$=0.1,0.01,0.001$. 
{\it Lower panel:} Data for the two seasons
folded on the period
$P=16.38$\,days. 
 \label{swlomb} }
\end{figure*}

The SuperWASP light curve of the active star WASP-52 shows long term modulations. 
EulerCam data also show irregularities. 
We have analyzed the SuperWASP light curve to determine whether they are 
periodic, and possibly due to the combination of  magnetic activity and the rotation of
the star.  
The value of  \vsini\ derived above together
with the estimated stellar radius implies a rotation period of 5\,--\,15~days,
assuming that the rotation axis of the star is approximately aligned with the
orbital axis of the planet. We show below in Sect.~\ref{sect_fit_RM}  that 
this is not exactly the case, but the misalignment remains low.
We used the sine-wave fitting method described in Maxted et 
al.~(\cite{maxted11}) to calculate the periodograms shown in Fig.~\ref{swlomb}.
These are calculated over 4096 uniformly spaced frequencies from 0 to 2.5
cycles/day. The false-alarm probability levels shown in these figures are
calculated using a boot-strap Monte Carlo method also described in 
Maxted et al.~(\cite{maxted11}). Variability due to star spots is not expected to be 
coherent on long timescales as a consequence of the finite lifetime of star-spots and
differential rotation in the photosphere, and so we analyzed the two seasons of
data for WASP-52 separately. 

The second season of data shows a highly significant periodic modulation with
a period of $16.4\pm0.04$ days and an amplitude of 9.6 milli-magnitudes. 
We used a boot-strap Monte Carlo method to estimate the error on this period. We
also calculated periodograms of three stars with similar magnitudes and colors to
WASP-52 observed simultaneously with the same WASP camera. None of these stars
showed significant periodic modulation in the same frequency interval.  The
strongest peak in the first season of data is marginally significant and is
consistent with the 1-day alias of the period detected in the second season of data. 

\begin{table*}
\caption[]{Planetary and stellar parameters for the systems WASP-52, 58, 59, and 60.}
\label{tab:mcmc}
\centering
\begin{centering}
\begin{tabular}{lcrrl}
\hline\\
Parameter & Symbol & WASP-52 & WASP-58 & Units \\
\hline\\
Transit epoch (HJD-2450000.0)     & $T_0$                 & $ 5793.68143 \pm    0.00009 $  & $ 5183.9335  \pm    0.0010  $  & days                  \\           
Orbital period                    & $P$                   & $ 1.7497798  \pm 0.0000012  $  & $ 5.017180   \pm 0.000011   $  & days                  \\           
Transit duration                  & $t_T$                 & $ 0.0754     \pm 0.0005     $  & $ 0.1582     \pm 0.0043     $  & days                  \\           
Planet/star area ratio            & $(R_{\rm p}/R_*)^2$   & $ 0.0271     \pm 0.0004     $  & $ 0.0145     \pm 0.0010     $  &                       \\           
Impact parameter                  & $b$                   & $ 0.60       \pm 0.02       $  & $ 0.46       \pm 0.23       $  & $R_*$                 \\           
Scaled stellar radius             & $R_*/a$               & $ 0.1355     \pm 0.0020     $  & $ 0.097      \pm 0.011      $  &                       \\           
Stellar density                   & $\rho_*$              & $ 1.76       \pm 0.08       $  & $ 0.58       \pm 0.19       $  & $\rho_\odot$          \\           
Stellar surface gravity           & $\log g_*$            & $ 4.582      \pm 0.014      $  & $ 4.27       \pm 0.09       $  &  (cgs)                \\           
Stellar radius                    & $R_*$                 & $ 0.79       \pm 0.02       $  & $ 1.17       \pm 0.13       $  &  R$_\odot$            \\           
Stellar mass                      & $M_*$                 & $ 0.87       \pm 0.03       $  & $ 0.94       \pm 0.10       $  &  M$_\odot$            \\           
Semi major axis                	  & $a$                   & $ 0.0272     \pm 0.0003     $  & $ 0.0561     \pm 0.0020     $  &  AU                   \\           
Orbital inclination               & $i_p$                 & $ 85.35      \pm  0.20      $  & $ 87.4       \pm  1.5       $  & $^\circ$              \\           
Stellar reflex velocity           & $K_1$                 & $ 0.0843     \pm 0.0030     $  & $ 0.1101     \pm 0.0041     $  & kms s$^{-1}$          \\           
Orbital eccentricity              & $e$                   &  0 (fixed)                     &  0 (fixed)                     &                       \\    
Planet radius                     & $R_{\rm p}$           & $ 1.27       \pm 0.03       $  & $ 1.37       \pm 0.20       $  &  R$_{\rm J}$          \\           
Planet mass                       & $M_{\rm p}$           & $ 0.46       \pm 0.02       $  & $ 0.89       \pm 0.07       $  &  M$_{\rm J}$          \\           
Planet surface gravity            & $\log g_{\rm p}$      & $ 2.81       \pm 0.03       $  & $ 3.03       \pm 0.12       $  &  (cgs)                \\           
Planet density                    & $\rho_{\rm p}$        & $ 0.22       \pm 0.02       $  & $ 0.34       \pm 0.14       $  & $\rho_{\rm J}$        \\           
Planetary equilibrium temperature & $T_{\rm P}$           & $ 1315       \pm   35       $  & $ 1270       \pm   80       $  & K                     \\           
\hline\\
Parameter & Symbol & WASP-59 & WASP-60 & Units \\
\hline\\
Transit epoch (HJD-2450000.0)     & $T_0$                 & $ 5830.95559 \pm    0.00053 $  & $ 5747.0295  \pm    0.0022  $  & days                  \\           
Orbital period                    & $P$                   & $ 7.919585   \pm 0.000010   $  & $ 4.3050011  \pm 0.0000062  $  & days                  \\           
Transit duration                  & $t_T$                 & $ 0.102      \pm 0.002      $  & $ 0.139      \pm 0.005      $  & days                  \\           
Planet/star area ratio            & $(R_{\rm p}/R_*)^2$   & $ 0.0169     \pm 0.0008     $  & $ 0.0060     \pm 0.0006     $  &                       \\           
Impact parameter                  & $b$                   & $ 0.29       \pm 0.18       $  & $ 0.37       \pm 0.23       $  & $R_*$                 \\           
Scaled stellar radius             & $R_*/a$               & $ 0.041      \pm 0.003      $  & $ 0.100      \pm 0.011      $  &                       \\           
Stellar density                   & $\rho_*$              & $ 3.11       \pm 0.64       $  & $ 0.72       \pm 0.20       $  & $\rho_\odot$          \\           
Stellar surface gravity           & $\log g_*$            & $ 4.72       \pm 0.06       $  & $ 4.35       \pm 0.09       $  &  (cgs)                \\           
Stellar radius                    & $R_*$                 & $ 0.613      \pm 0.044      $  & $ 1.14       \pm 0.13       $  &  R$_\odot$            \\           
Stellar mass                      & $M_*$                 & $ 0.719      \pm 0.035      $  & $ 1.078      \pm 0.035      $  &  M$_\odot$            \\           
Semi major axis                   & $a$                   & $ 0.0697     \pm 0.0011     $  & $ 0.0531     \pm 0.0006     $  &  AU                   \\           
Orbital inclination               & $i_p$                 & $ 89.27      \pm  0.52      $  & $ 87.9       \pm  1.6       $  & $^\circ$              \\           
Stellar reflex velocity           & $K_1$                 & $ 0.1096     \pm 0.0044     $  & $ 0.0608     \pm 0.0038     $  & kms s$^{-1}$          \\           
Orbital eccentricity              & $e$                   & $ 0.100      \pm 0.042      $  &  0 (fixed)                     &                       \\           
Argument of periastron            & $\omega$              & $ 74         \pm 15         $  &      --                        & $^\circ$              \\           
Planet radius                     & $R_{\rm p}$           & $ 0.775      \pm 0.068      $  & $ 0.86       \pm 0.12       $  &  R$_{\rm J}$          \\           
Planet mass                       & $M_{\rm p}$           & $ 0.863      \pm 0.045      $  & $ 0.514      \pm 0.034      $  &  M$_{\rm J}$          \\           
Planet surface gravity            & $\log g_{\rm p}$      & $ 3.52       \pm 0.08       $  & $ 3.19       \pm 0.12       $  &  (cgs)                \\           
Planet density                    & $\rho_{\rm p}$        & $ 1.8        \pm 0.5        $  & $ 0.8        \pm 0.3        $  & $\rho_{\rm J}$        \\           
Planetary equilibrium temperature & $T_{\rm P}$           & $ 670        \pm  35        $  & $ 1320       \pm   75       $  & K                     \\           
\hline\\
 \label{table_parameters_fit}
\end{tabular}
\end{centering}
\end{table*}

\subsection{Parameters of the planetary systems}

\subsubsection{Method}

For a given system, all the available data were fitted simultaneously: 
the radial velocities with a Keplerian orbit, and the transit light curves with the 
Mandel \& Agol~(\cite{mandel02}) algorithm using four non-linear limb-darkening coefficients 
from the tabulations of Claret~(\cite{claret04}). 
The global fit was done with the latest version of the Markov-chain Monte-Carlo (MCMC) 
procedure presented in detail by Collier Cameron et al.~(\cite{cameron07b}) and 
Pollacco et al.~(\cite{pollacco08}). 
The free parameters in the fits are 
the transit epoch $T_0$, 
the orbital period $P$, 
the depth of the transit or the planet/star area ratio $(R_{\rm p}/R_*)^2$, 
the total transit duration $t_T$, 
the impact parameter $b$,
the eccentricity $e$ of the orbit, 
the argument of periastron $\omega$,
the semi-amplitude $K_1$ of the radial velocity variations,
the $n$ systemic radial velocities $\gamma_n$ (one for each of the $n$ radial velocity datasets).
When no significant eccentricity is detected (in three out of the four systems), the 
orbits are considered as circular in the final fit.
The stellar density $\rho_*$ is directly derived from the transit light curve 
(through $P$, $t_T$, and $b$ values). The density, together with \teff\  and [Fe/H], yield  the stellar mass $M_*$ and radius $R_*$ 
through empirical calibrations established for eclipsing binaries (Torres et al.~\cite{torres10}; Enoch et al.~\cite{enoch11}).
From these we conclude that the four host targets are dwarf stars.
The planetary mass $M_{\rm p}$ and radius $R_{\rm p}$ are derived from $K_1$  
and $(R_{\rm p}/R_*)^2$, which directly provides the planetary density  $\rho_{\rm p}$.
We also compute the derived parameters as 
the surface gravity $\log g_*$ (which here is more accurate than 
that obtained from spectral analysis), the semi major axis $a$ and the orbital inclination $i_p$, as well as 
the equilibrium temperature $T_{\rm P}$ of the planet, assuming it to be 
a black body and that energy~is efficiently redistributed from the planetÕs day-side to its night-side. 

\begin{figure*}[t!] 
\begin{center}
\includegraphics[scale=0.9]{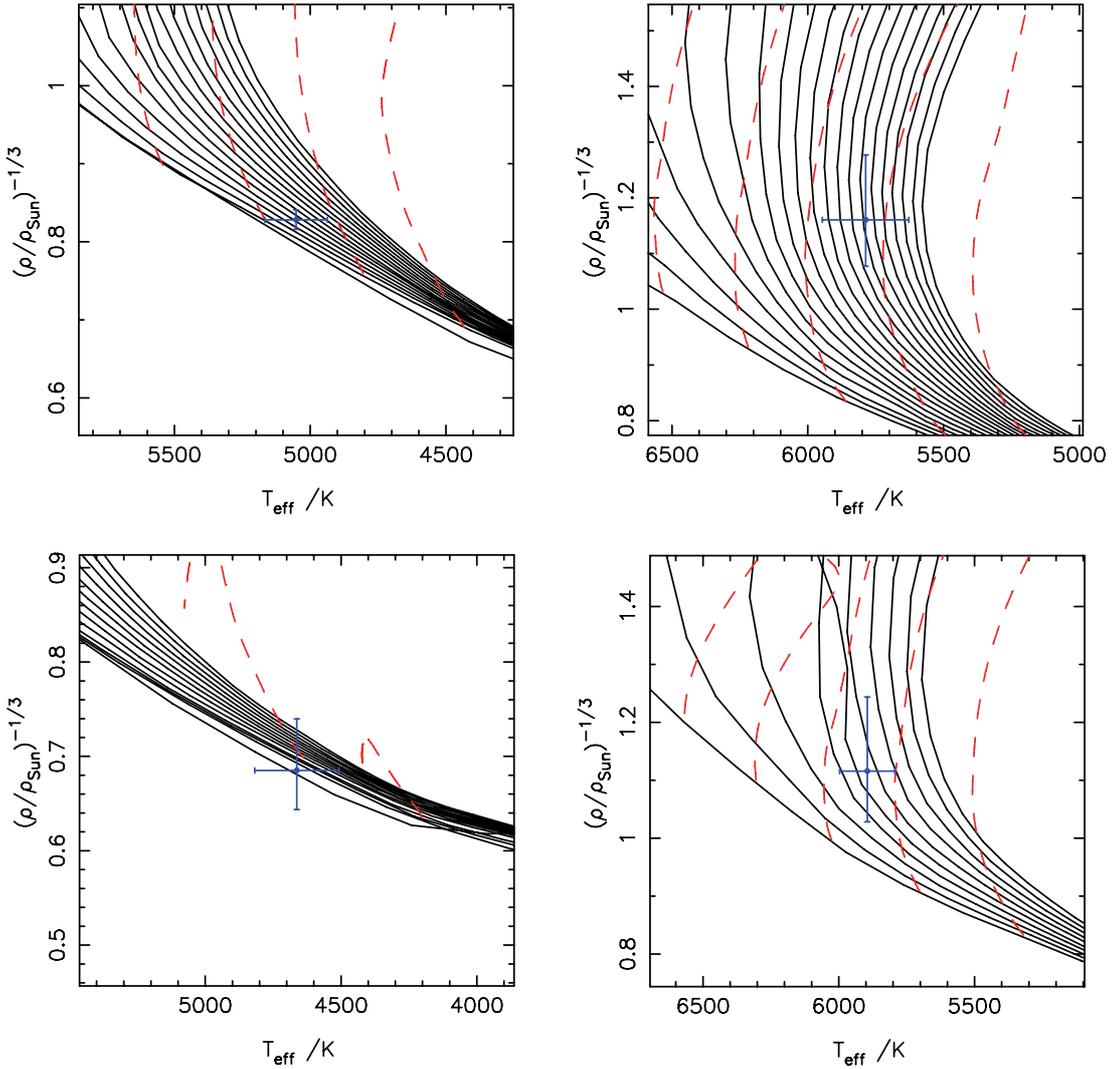}
\caption{HR diagrams for WASP-52, 58, 59, and 60 (Yonsei-Yale stellar models). 
Each host star is denoted by a blue cross as a function of the stellar density and 
the effective temperature. 
Solid, black lines are the isochrones, and red, dashed lines are the evolutioanry tracks
for different masses.
WASP-52 (upper, left):  0.1\,Gyr isochrone, and then 1\,Gyr to 15\,Gyr in 1\,Gyr increments,
and mass tracks from 1.0 to 0.7\,\MS\ in 0.1 increments from left to right.
WASP-58 (upper, right): 1 to 20\,Gyr in 1\,Gyr increment isochrones, and mass tracks from 1.1 to 0.7\,\MS\ 
in 0.1 increments from left to right.
WASP-59 (lower, left): 0.1, 0.4, 0.6\,Gyr isochrones, then 1\,Gyr to 10\,Gyr in 1\,Gyr increments, 
and 0.7 and 0.6\,\MS\  mass tracks from left to right.
WASP-60 (lower, right): 1\,Gyr to 10\,Gyr in 1\,Gyr increments,
and mass tracks from 1.3 to 0.70\,\MS\ in 0.1 increments from left to right.
}
\label{fig_track}
\end{center}
\end{figure*}

For comparison, the stellar mass is also derived from mass tracks in $\rho_*$-\textit{vs.}-\teff\ 
HR diagrams (Sozzetti et al.~\cite{sozzetti07}). We compared several 
models for the tracks in the HR diagrams:
the Padova isochrones of Marigo et al.~(\cite{marigo10}), the Teramo stellar models from 
Pietrinferni et al.~(\cite{pietrinferni04}), the Victoria-Regina stellar models from 
Vandenberg et al.~(\cite{vandenberg06}), and the Yonsei-Yale isochrones of 
Demarque et al.~(\cite{demarque04}).
The last among these are plotted in Fig.~\ref{fig_track} as examples.
We also estimate the stellar ages from the isochrones in these HR diagrams; although their uncertainties are of order several Gyr, they remain 
in broad agreement with gyrochronological~ages. 

The best fits of the WASP-52, 58, 59, and 60 systems are plotted over the data in 
Figs.~\ref{fig_obs_W52}, \ref{fig_obs_W58}, \ref{fig_obs_W59}, and \ref{fig_obs_W60}.
The median values and 1-$\sigma$ uncertainties of the parameters obtained from 
the MCMC are reported in Table~\ref{table_parameters_fit}. 
Each system is detailed below.

\subsubsection{WASP-52}

There is a good agreement between the stellar mass obtained from the MCMC and 
that estimated from mass tracks in evolution diagrams for WASP-52. 
The red noise levels in the light curves are negligible; averaging the photometric data
in 6- to 10-minute bins made no appreciable difference to the error bars on the derived parameters.
The radial velocity fit provide systemic radial velocities of
$-0.8953 \pm 0.0075$,
$-1.0415 \pm 0.0033$,
$-0.9265 \pm 0.0062$, and
$-0.8877 \pm 0.0040$\,k\ms\ 
for SOPHIE HE1, HE2, HR, and CORALIE, respectively, with a dispersion 
of the residuals of the order of 13\,\ms. 
We fit the data using only parts of the radial velocity datasets, which provide 
similar results. This suggests that the systematic shift of $146 \pm 8$\,\ms\  between 
the two SOPHIE HE datasets (see Sect.~\ref{sect_sophie}) has no significant 
impact on the derived parameters of the system. 
There is a hint of a drift of $\sim+40$\,\ms\ over the 15-month time span of our observations.
It could be due to an extra component in the system, but the unknown systematic 
shifts between datasets do not allow any firm conclusion to be drawn on the source of the drift.
No significant eccentricity is~detected.

The orbital period is 1.75\,day. The one-day alias of that signal is 2.33\,days; it should be 
considered as the photometric and radial velocity observations were made  with a typical 
1-day frequency. 
The periodograms show signal at both frequencies 1.75 and 2.33\,days. However, the radial velocity data 
folded at 2.33\,days clearly  provide a poorer fit, with a dispersion of 30\,\ms\ in the residuals, 
more than two times larger than the 1.75-day solution. 
This confirms that 1.75\,day is the correct orbital period of the planet.

WASP-52b is a hot jupiter which is less than half the mass of Jupiter, but with a significantly larger radius.
It is thus a new example of an inflated hot jupiter. Together with the small radius of its 
K2V host star, it makes the planetary transits particularly deep at 2.7\,\%. 
The case of WASP-52b should act as a warning, since many transit signals of this large 
depth detected by photometric surveys may be flagged as due to binaries, and a priori 
removed from radial-velocity follow-up (see also Southworth et al.~\cite{southworth12}). 
Hence, there may be an underlying population 
of heavily inflated planets with large transit depths that have been mistakenly identified 
as binaries and hence their true nature has never been revealed.

\subsubsection{WASP-58}

WASP-58 has larger uncertainties on its derived parameters as no extra transit light curves 
were obtained in complement to the original SuperWASP data.
Assuming 
a main-sequence mass-radius prior would reduce the uncertainties, in particular on the 
stellar and planetary radii, but we did not make that assumption in our final results in 
order to remain conservative.  This does not change the parameters significantly.

For that target, the empirical estimation of the stellar mass derived from 
\teff\  and [Fe/H] (Torres et al.~\cite{torres10}; Enoch et al.~\cite{enoch11}) 
provides $M_* = 1.03 \pm 0.05$\,\MS, whereas the isochrones favour lower 
values around $M_* = 0.82 \pm 0.06$\,\MS. This discrepancy may be due 
to the low metallicity of the star, which could be beyond  the validity of the 
Torres-Enoch calibration. Metal-poor main-sequence stars of
a given mass do indeed burn hotter than their metal-rich counterparts. 
The metallicity of WASP-58 has to be increased to 
at least $-0.2$ to allow a better agreement between the two estimations. 
However, when we consider other cases where hot jupiters orbit low-metallicity 
host stars (e.g. WASP-21, 37, or 78) we do not find similar discrepancies between 
the two estimates. Thus we remain uncertain of the true source of the disagreement. 
To reflect this, we have conservatively adopted an intermediate value for the 
stellar mass with an increased uncertainty to $M_* = 0.94 \pm 0.10$\,\MS. 

The radial velocities have a systemic value  of  $-28.6631 \pm 0.0033$\,k\ms\ 
and their residuals show a 11-\ms\ dispersion, in agreement with the expected uncertainties
on their measurements. No structure or drift are seen in
the residuals, providing no evidence of any extra component in the~system. 

Regardless of whether we assume or not a mass-radius relationship for the host star, 
we find that WASP-58b is
a typical inflated hot jupiter, with a radius 1.4~times that of Jupiter, 
and a similar mass. Its orbital period of 5.0\,days is located in the higher tail of the hot-jupiter period distribution. It orbits a particularly metal-poor star, whereas most hot jupiters preferentially orbit metal-rich~stars.

\subsubsection{WASP-59}

The systemic radial velocity of WASP-59 is $-56.7976 \pm 0.0043$\,k\ms\, with a 
10-\ms\ dispersion around the circular Keplerian fit and no detection of any 
additional drift. 
The stellar mass estimate from the MCMC analysis agrees with that derived from 
evolutionary tracks in the density-temperature plane. 
As expected for late K-type stars, the isochrones are too closely-spaced to allow them to be 
distinguished, and gyrochronology provides the only age estimate. 
By allowing the eccentricity to vary freely, 
we found $e=0.10\pm 0.04$ and a dispersion of the radial velocity residuals 
decreased to 5\,\ms. 
If the marginal (2\,$\sigma$ ) detection of an eccentric orbit is confirmed, then it may be 
in part due to the 7.9-day orbital period, which is relatively long compared to typical 
hot jupiters. Indeed,  tidal circularization is 
less effective at larger orbital separations, and initial eccentricities in the planetary 
orbit may not yet have been damped
(e.g. Hut~\cite{hut81}). Following Matsumura et al.~(\cite{matsumura08}), 
we estimated the circularization time to be 5\,Gyr, which is about ten times longer than 
the gyrochronological age. 

WASP-59b is one of the rare hot jupiters found by ground-based transit 
searches with orbital periods longer than seven 
days, together with WASP-8b, HAT-P-15b, and HAT-P-17b. 
Detecting such transiting objects from ground-based photometry from a single site remains difficult
because of the night/day duty cycle. 
The space-based missions CoRoT and Kepler provide continuous photometry, 
and are therefore more suited to detect planets with longer orbital periods such as
CoRoT-4b, 6b, 10b, 20b, and Kepler-28c, 29b, 29c, 32c.

The 0.7-\RJ\ radius of WASP-59b places it just between the hot jupiters (with radii typically 
between 1.0 and 1.4\,\RJ\ and periods below 10 days) and the low-radius planets recently detected
mainly by Kepler (with radii typically between 0.1 and 0.5\,\RJ). Only a few objects 
are known in that intermediate radius range. One of the few others is the planet HD\,149026b (Sato~et al.~\cite{sato05}), whose density indicates a large 
dense core . 
Interestingly enough, the planet WASP-59b also has a particularly high density.
Its host star is also one of the rare stars cooler than 5000\,K that harbour a hot jupiter.

\subsubsection{WASP-60}

The systemic radial velocities of WASP-60 are $-26.4967 \pm 0.0103$\,k\ms\ and
$-26.6288 \pm 0.0054$\,k\ms\ for the two HE datasets. The systematic shift 
between them ($132 \pm 12$\,\ms) agrees with that found above for WASP-52
($146 \pm 8$\,\ms); this reinforces our conclusion that this shift has an instrumental 
origin (see Sect.~\ref{sect_sophie}). The radial velocity residuals show a 20-\ms\  
dispersion, in agreement with the expected error bars on their measurements.
There is a hint of a drift of $\sim+70$\,\ms\  over the 16-month time span of our 
observations. Again, this may suggest the presence of an additional body 
in the system, but the unknown systematic shift between the two datasets 
prevents any firm conclusion from being drawn.

The stellar mass obtained from the MCMC analysis agrees with that estimated
from evolutionary tracks in the density-temperature plane diagrams for WASP-60. 
The planet appears to have a Saturn-like radius and sub-Jupiter mass, making it 
an interestingly compact for a planet orbiting a  slightly metal-poor star.

\subsection{Rossiter-McLaughlin anomaly analysis of WASP-52}   
\label{sect_fit_RM}

The radial velocities of WASP-52 measured during the 2011 August 21 transit 
were fitted in order to derive the sky-projected angle $\lambda$ between 
the planetary orbital axis and the stellar rotation axis. 
We applied the analytical approach developed by Ohta et al.~(\cite{otha05})
to model the form of the Rossiter-McLaughlin anomaly, which is described by 
ten parameters: 
the stellar limb-darkening linear coefficient $\epsilon$, 
the transit parameters $R_\mathrm{p}/R_*$, $a/R_*$ and $i_p$, 
the parameters of the circular orbit ($P$, $T_0$, and $K_1$), 
the systemic radial velocity,
and finally \vsini\ and $\lambda$.
We adopted $\epsilon=0.82$ computed by Claret~(\cite{claret04})
for the $g^\prime$ filter corresponding to the SOPHIE 
wavelength range. The transit and orbital parameters were determined above  
from the light curves and radial velocity fits. Their uncertainties are negligible here
for the fit of the Rossiter-McLaughlin anomaly.
We checked that varying them in their 1-$\sigma$ intervals 
has no impact on the quality of the fit nor on the derived $\lambda$ value.
The main parameters that play a role in 
this fit are the systemic velocity of that particular dataset, $\lambda$, and \vsini. 
As these parameters could be correlated in the Rossiter-McLaughlin fit, we computed 
the \kid\ of the fit on a three-dimensional grid scanning all their possible values.

\begin{figure}[ht!] 
\begin{center}
\includegraphics[scale=0.53]{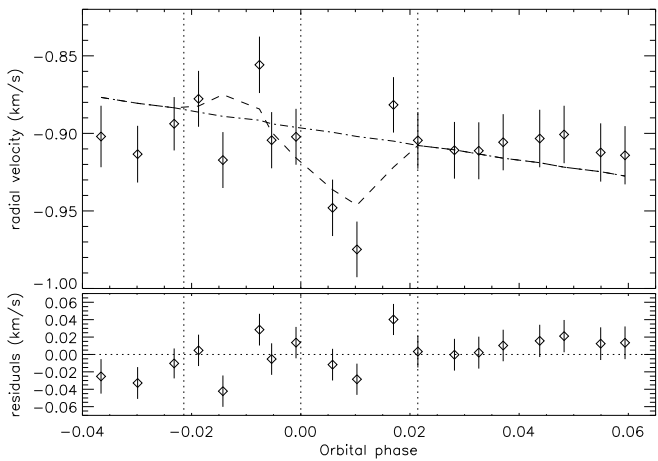}
\caption{Spectroscopic observation of the transit of WASP-52b. 
\textit{Top:} SOPHIE radial velocity measurements as a function of 
the orbital phase (diamonds), Keplerian fit ignoring the transit 
(dotted-dashed line), and final fit including the model of the Rossiter-McLaughlin 
anomaly (dashed line).
The vertical dotted lines show the times of mid-transit, first, 
and fourth contacts.  
\textit{Bottom:} Residuals of the final fit.}
\label{fig_RM_W52}
\end{center}
\end{figure}

The systemic velocity was determined with a $\pm5$-\ms\  accuracy. 
It is constrained thanks to the observations secured immediately before and 
after the transit.
The confidence interval contours estimated from \kid\  variations
for the $\lambda$ and \vsini\ show no correlation between these two parameters, 
as expected for systems with impact parameter $b$ significantly different from zero.
From \kid\ variations,  we obtained $\lambda = 24^{\circ}$$^{+17}_{-9}$ and 
\vsini$\, = 2.5 \pm 1.0$\,\kms. The best fit is plotted in Fig.~\ref{fig_RM_W52}. 

The dispersion of the residuals about the fit is 20\,\ms, in agreement with the expected uncertainties.
The Rossiter-McLaughlin anomaly  detection is noisy and only spans a few points.
As a statistical test for the detection, 
we computed the \kid\ over the measurements secured during the transit night 
for the fits 
first including the Rossiter-Mclaughlin anomaly, and then compared this to the fits 
achieved when the RM anomaly is ignored.
The F-test indicates that the \kid\ improvement with the Rossiter-McLaughlin fit
has a probability $>65$\% to be actually due to the Rossiter-McLaughlin anomaly detection.
We  therefore concluded that the spectroscopic transit is significantly detected.

Due to the quality of the the data, we remain cautious with respect to the numerical result
on $\lambda$.
The orbit of WASP-52b is apparently prograde and slightly misaligned.
We note that  the \vsini\ value obtained from the fit agrees with the value obtained 
from spectral analysis (see Table~\ref{table_stell_params}).

\section{Conclusion}
\label{sect_concle}

We have presented the discovery of four new transiting exoplanets, 
WASP-52b, WASP-58b, WASP-59b, and WASP-60b, 
detected mainly from SuperWASP-North and SOPHIE observations. As with most of the transiting planets found 
from ground-based observatories, they can be ranked in the hot-jupiter class. Each of them however 
has some peculiarities. WASP-52b and WASP-58b are new cases of inflated planets, increasing the
population of giant planets with abnormally large radii whose origin is not well understood yet despite 
the numerous models proposed to explain them (see, e.g., Fortney \&\ Nettelmann~\cite{fortney10}).
WASP-52 is a new case of a possibly slightly misaligned system; it does not fit however with the 
possible correlations of $\lambda$ with \teff\ (Winn et al.~\cite{winn10}) or $M_p$ (H\'ebrard et 
al.~ \cite{hebrard11}). The hot jupiter WASP-58b orbits a particularly metal-poor star.
WASP-59b has an orbital period longer than typical transiting hot jupiters, and a high density, 
suggesting a large core. Finally, WASP-60b presents shallow transits of 0.6\,\%, illustrating 
the capabilities of ground-based surveys to detect small-size planets. These new planets 
provide new targets for follow-up studies and increase 
our statistics on well-characterized exoplanets, allowing progress 
to be made in the comprehension of the formation and evolution of planetary systems.

\begin{acknowledgements}
We thank the 2010 cohort of the AS4025 Observational Astrophysics class at the University of 
St Andrews for their help in analyzing and eliminating astrophysical false 
positives in the JGT data, and for recognising WASP-52 as a good potential 
planet candidate.
The research leading to these results has received funding from the European 
CommunityÕs Seventh Framework Programme (FP7/2007-2013) under grant 
agreement number RG226604 (OPTICON), 
as well as the Programme National de Plan\'etologie (PNP) of CNRS/INSU (France).
RFD is supported by CNES.
\end{acknowledgements}

\begin{table*}
  \caption{Radial velocities (electronic table).}
\begin{tabular}{lccrr}
\hline
\hline
HJD$_{\rm UTC}$ & RV & $\pm$$1\,\sigma$ & Bis. span & Instrument/ \\
-2\,400\,000 & (km\,s$^{-1}$) & (km\,s$^{-1}$) & (km\,s$^{-1}$) & mode \\
\hline
\multicolumn{3}{l}{\textbf{\hspace{0.7cm}WASP-52}}  \\
55485.3531$^\dagger$	&	-0.821	&	0.011	&	-0.039	& SOPHIE HE1 \\
55486.4051$^\dagger$	&	-0.953	&	0.014	&	-0.049	& SOPHIE HE1 \\
55488.3582$^\dagger$	&	-0.883	&	0.013	&	-0.028	& SOPHIE HE1 \\
55489.3710$^\dagger$	&	-0.934	&	0.016	&	-0.025	& SOPHIE HE1 \\
55519.3509$^\dagger$	&	  -1.127	&	0.012	&	-0.035	& SOPHIE HE2 \\
55523.2950$^\dagger$	&	  -1.079	&	0.011	&	-0.017	& SOPHIE HE2 \\
55524.3205$^\dagger$	&	  -1.104	&	0.015	&	 0.009	& SOPHIE HE2 \\
55525.3292$^\dagger$	&	  -1.001	&	0.011	&	-0.047	& SOPHIE HE2 \\
55527.3109			&	  -0.963	&	0.012	&	 0.007	& SOPHIE HE2 \\
55529.3862			&	  -1.010	&	0.010	&	-0.021	& SOPHIE HE2 \\
55540.3215			&	  -1.145	&	0.018	&	-0.016	& SOPHIE HE2 \\
55559.2497			&	  -1.052	&	0.010	&	-0.021	& SOPHIE HE2 \\
55563.2598			&	  -1.100	&	0.010	&	-0.017	& SOPHIE HE2 \\
55564.2396			&	  -0.969	&	0.011	&	-0.004	& SOPHIE HE2 \\
55579.2475$^\dagger$	&	  -1.075	&	0.012	&	-0.035	& SOPHIE HE2 \\
55581.2370$^\dagger$	&	  -0.994	&	0.012	&	-0.011	& SOPHIE HE2 \\
55583.2417$^\dagger$	&	  -0.935	&	0.015	&	 0.011	& SOPHIE HE2 \\
55585.2437$^\dagger$	&	  -0.966	&	0.011	&	-0.044	& SOPHIE HE2 \\
55586.2770$^\dagger$	&	  -1.068	&	0.010	&	-0.015	& SOPHIE HE2 \\
55588.2466$^\dagger$	&	  -0.983	&	0.010	&	 0.011	& SOPHIE HE2 \\
55745.5740			&	-0.935	&	0.014	&	-0.030	& SOPHIE HR \\
55753.6075			&	-0.982	&	0.018	&	-0.021	& SOPHIE HR \\
55760.6012$^\dagger$	&	-0.984	&	0.031	&	-0.067	& SOPHIE HR \\
55763.5721$^\dagger$	&	-0.850	&	0.020	&	-0.018	& SOPHIE HR \\
55764.5729$^\dagger$	&	-0.979	&	0.023	&	-0.006	& SOPHIE HR \\
55766.5778			&	-0.916	&	0.019	&	-0.086	& SOPHIE HR \\
55768.5387			&	-0.886	&	0.015	&	 0.021	& SOPHIE HR \\
55771.5908			&	-0.982	&	0.012	&	 0.010	& SOPHIE HR \\
55773.5791			&	-0.914	&	0.010	&	-0.027	& SOPHIE HR \\
55413.7375			&	-0.863	&	0.015	&	0.028	& CORALIE \\
55485.6031			&	-0.880	&	0.022	&	-0.008	& CORALIE \\	
55538.5713			&	-0.996	&	0.020	&	0.013	& CORALIE \\
55541.5460			&	-0.829	&	0.018	&	0.091	& CORALIE \\
55725.8837			&	-0.965	&	0.011	&	-0.002	& CORALIE \\
55726.9038			&	-0.796	&	0.011	&	0.022	& CORALIE \\
55756.8692			&	-0.837	&	0.019	&	0.006	& CORALIE \\
55769.8876			&	-0.934	&	0.014	&	0.018	& CORALIE \\
55777.8219			&	-0.846	&	0.013	&	0.052	& CORALIE \\
55782.7876			&	-0.804	&	0.013	&	0.098	& CORALIE \\
55784.8728			&	-0.868	&	0.013	&	0.019	& CORALIE \\
55795.7326			&	-0.962	&	0.013	&	0.008	& CORALIE \\
55804.6863			&	-0.962	&	0.015	&	0.028	& CORALIE \\
55829.7201			&	-0.851	&	0.010	&	0.018	& CORALIE \\
55832.6556			&	-0.966	&	0.012	&	0.026	& CORALIE \\
55858.5814			&	-0.933	&	0.023	&	0.095	& CORALIE \\
55795.3677			&	-0.902	&	0.020	&	 ---		& SOPHIE HE-RM \\
55795.3802			&	-0.913	&	0.018	&	 ---		& SOPHIE HE-RM \\
55795.3900			&	-0.894	&	0.017	&	 ---		& SOPHIE HE-RM \\
55795.3991			&	-0.878	&	0.018	&	 ---		& SOPHIE HE-RM \\
55795.4081			&	-0.917	&	0.018	&	 ---		& SOPHIE HE-RM \\
55795.4163			&	-0.856	&	0.018	&	 ---		& SOPHIE HE-RM \\
55795.4236			&	-0.904	&	0.018	&	 ---		& SOPHIE HE-RM \\
55795.4313			&	-0.902	&	0.018	&	 ---		& SOPHIE HE-RM \\
55795.4417			&	-0.948	&	0.018	&	 ---		& SOPHIE HE-RM \\
55795.4506			&	-0.975	&	0.018	&	 ---		& SOPHIE HE-RM \\
55795.4604			&	-0.882	&	0.018	&	 ---		& SOPHIE HE-RM \\
55795.4694			&	-0.904	&	0.018	&	 ---		& SOPHIE HE-RM \\
55795.4790			&	-0.911	&	0.018	&	 ---		& SOPHIE HE-RM \\
55795.4878			&	-0.911	&	0.018	&	 ---		& SOPHIE HE-RM \\
55795.4969			&	-0.906	&	0.018	&	 ---		& SOPHIE HE-RM \\
55795.5065			&	-0.903	&	0.019	&	 ---		& SOPHIE HE-RM \\
55795.5164			&	-0.901	&	0.019	&	 ---		& SOPHIE HE-RM \\
55795.5264			&	-0.912	&	0.019	&	 ---		& SOPHIE HE-RM \\
55795.5364			&	-0.914  	& 	0.019	&	 ---		& SOPHIE HE-RM \\
\hline
\hline
\multicolumn{5}{l}{$\dagger$: measurements corrected for moonlight pollution.} \\
\end{tabular}
\end{table*}
\begin{table*}
\begin{tabular}{lccrr}
\hline
\hline
HJD$_{\rm UTC}$ & RV & $\pm$$1\,\sigma$ & Bis. span & Instrument/ \\
-2\,455\,000 & (km\,s$^{-1}$) & (km\,s$^{-1}$) & (km\,s$^{-1}$) & mode \\
\hline
\multicolumn{3}{l}{\textbf{\hspace{0.7cm}WASP-58}}  \\
55744.4674			&	-28.558	&	0.012	&	-0.014	& SOPHIE HE \\
55746.4222			&	-28.744	&	0.011	&	-0.023	& SOPHIE HE \\
55752.4138$^\dagger$	&	-28.758	&	0.013	&	-0.032	& SOPHIE HE \\
55753.3777$^\dagger$	&	-28.660	&	0.016	&	-0.038	& SOPHIE HE \\
55754.3663$^\dagger$	&	-28.523	&	0.014	&	-0.010	& SOPHIE HE \\
55763.4573$^\dagger$	&	-28.672	&	0.013	&	-0.038	& SOPHIE HE \\
55764.3750			&	-28.575	&	0.011	&	-0.046	& SOPHIE HE \\
55765.4067			&	-28.593	&	0.011	&	-0.018	& SOPHIE HE \\
55766.3448			&	-28.720	&	0.015	&	 0.011	& SOPHIE HE \\
55767.3736			&	-28.767	&	0.014	&	-0.059	& SOPHIE HE \\
55768.3734			&	-28.672	&	0.011	&	-0.015	& SOPHIE HE \\
55772.4158			&	-28.765	&	0.010	&	-0.033	& SOPHIE HE \\
55774.3833			&	-28.560	&	0.009	&	-0.020	& SOPHIE HE \\
55776.4381			&	-28.734	&	0.011	&	-0.028	& SOPHIE HE \\
55777.3450$^\dagger$	&	-28.761	&	0.014	&	-0.018	& SOPHIE HE \\
\hline
\multicolumn{3}{l}{\textbf{\hspace{0.7cm}WASP-59}}  \\
55745.5453			&	-56.897	&	0.014	&	-0.007		& SOPHIE HE \\
55757.5430$^\dagger$	&	-56.688	&	0.028	&	-0.004		& SOPHIE HE \\
55763.5922$^\dagger$	&	-56.800	&	0.013	&	-0.007		& SOPHIE HE \\
55764.6287$^\dagger$	&	-56.726	&	0.015	&	-0.026		& SOPHIE HE \\
55766.5375$^\dagger$	&	-56.701	&	0.014	&	 0.004		& SOPHIE HE \\
55768.5633			&	-56.883	&	0.010	&	-0.034		& SOPHIE HE \\
55771.5172			&	-56.800	&	0.010	&	 0.026		& SOPHIE HE \\
55772.5616			&	-56.748	&	0.010	&	-0.014		& SOPHIE HE \\
55773.5297			&	-56.694	&	0.009	&	 0.010		& SOPHIE HE \\
55774.5097			&	-56.699	&	0.008	&	 0.006		& SOPHIE HE \\
55776.5272			&	-56.873	&	0.010	&	-0.027		& SOPHIE HE \\
55777.5466			&	-56.918	&	0.010	&	-0.029		& SOPHIE HE \\
\hline
\multicolumn{3}{l}{\textbf{\hspace{0.7cm}WASP-60}}  \\
55141.4975$^\dagger$	&	-26.530	&	0.014	&	0.001	& SOPHIE HE1 \\
55142.5016$^\dagger$	&	-26.459	&	0.014	&	0.016	& SOPHIE HE1 \\
55485.4274$^\dagger$	&	-26.548	&	0.014	&	0.014	& SOPHIE HE1 \\
55486.4701$^\dagger$	&	-26.512	&	0.017	&	-0.024	& SOPHIE HE1 \\
55487.4957$^\dagger$	&	-26.433	&	0.017	&	-0.039	& SOPHIE HE1 \\
55488.3192$^\dagger$	&	-26.476	&	0.018	&	0.050	& SOPHIE HE1 \\
55489.4001$^\dagger$	&	-26.564	&	0.016	&	0.010	& SOPHIE HE1 \\
55524.3454$^\dagger$	&	-26.713	&	0.015	&	0.030	& SOPHIE HE2 \\
55525.3509$^\dagger$	&	-26.656	&	0.015	&	0.005	& SOPHIE HE2 \\
55529.4066			&	-26.666	&	0.013	&	0.001	& SOPHIE HE2 \\
55551.3639$^\dagger$	&	-26.642	&	0.017	&	0.074	& SOPHIE HE2 \\
55559.2666			&	-26.649	&	0.013	&	0.021	& SOPHIE HE2 \\
55563.2774			&	-26.682	&	0.013	&	0.016	& SOPHIE HE2 \\
55564.2611			&	-26.629	&	0.013	&	-0.024	& SOPHIE HE2 \\
55577.2737$^\dagger$	&	-26.620	&	0.014	&	0.009	& SOPHIE HE2 \\
55578.2567$^\dagger$	&	-26.572	&	0.014	&	0.035	& SOPHIE HE2 \\
55580.2448$^\dagger$	&	-26.686	&	0.014	&	0.007	& SOPHIE HE2 \\
55581.2597$^\dagger$	&	-26.662	&	0.015	&	0.021	& SOPHIE HE2 \\
55582.2619$^\dagger$	&	-26.597	&	0.027	&	0.002	& SOPHIE HE2 \\
55584.2465$^\dagger$	&	-26.692	&	0.015	&	0.020	& SOPHIE HE2 \\
55587.2500			&	-26.602	&	0.015	&	0.017	& SOPHIE HE2 \\
55597.2487			&	-26.663	&	0.014	&	-0.022	& SOPHIE HE2 \\
55597.2622			&	-26.681	&	0.013	&	-0.017	& SOPHIE HE2 \\
55598.2603			&	-26.635	&	0.013	&	0.034	& SOPHIE HE2 \\
55598.2723			&	-26.629	&	0.013	&	-0.010	& SOPHIE HE2 \\
55599.2508			&	-26.552	&	0.013	&	0.033	& SOPHIE HE2 \\
55599.2627			&	-26.532	&	0.012	&	0.019	& SOPHIE HE2 \\
55611.2832$^\dagger$	&	-26.689	&	0.028	&	0.030	& SOPHIE HE2 \\
55613.2682$^\dagger$	&	-26.613	&	0.023	&	-0.025	& SOPHIE HE2 \\
55614.2691$^\dagger$	&	-26.723	&	0.032	&	0.020	& SOPHIE HE2 \\
55615.2806			&	-26.657	&	0.020	&	-0.003	& SOPHIE HE2 \\
\hline
\multicolumn{5}{l}{$\dagger$: measurements corrected for moonlight pollution.} \\
  \label{table_rv}
\end{tabular}
\end{table*}


\begin{thebibliography}{}

\bibitem[2009]{baglin09} 
Baglin, A., Auvergne, M., Barge, P., et al. 2009, Transiting Planets, Proc. IAU
Symp., 253, 71

\bibitem[2007]{bakos07} 
Bakos, G. \`A., Noyes, R. W., Kov\`acs, G., et al. 2007, \apj, 656, 552 

\bibitem[1996]{baranne96} 
Baranne, A., Queloz, D., Mayor, M., et al. 1996, \aaps, 119, 373

\bibitem[Barnes (2007)]{2007ApJ...669.1167B} Barnes, S.A. 2007, \apj, 669, 1167

\bibitem[2011]{barros11} 
Barros, S. C. C., Faedi, F., Collier Cameron, A., et al. 2011, \aap, 525, A54

\bibitem[2010]{boisse10} 
Boisse, I., Eggenberger, A., Santos, N. C., et al. 2010, \aap, 523, A88

\bibitem[2010]{borucki10} 
Borucki, W. J., Koch, D. G., Basri, G., et al. 2010, Science, 327, 977

\bibitem[2009]{bouchy09} 
Bouchy, F., H\'ebrard, G., Udry, S., et al. 2009, \aap, 505, 853

\bibitem[2010]{bouchy10} 
Bouchy, F., Hebb, L., Skillen, I., et al. 2010,  \aap, 519, A98

\bibitem[2012]{bouchy12} 
Bouchy, F., D\'{\i}az, R.~F., H\'ebrard, G., et al. 2012, \aap, in press

\bibitem[Bruntt et al. 2010a]{2010A&A...519A..51B} 
Bruntt, H., Deleuil, M., Fridlund, M., et al. 2010a, \aap, 519, A51 

\bibitem[Bruntt et al. (2010b)]{2010MNRAS.405.1907B} 
Bruntt, H., Bedding, T.R., Quirion, P.-O., et al. 2010b, \mnras, 405, 1907

\bibitem[2004]{claret04} 
Claret, A. 2004, \aap, 428, 1001

\bibitem[2006]{cameron06} 
Collier Cameron, A., Pollacco, D., Street, R. A., et al. 2006, \mnras, 373, 799

\bibitem[2007a]{cameron07a} 
Collier Cameron, A., Bouchy, F., H\'ebrard, G., et al. 2007a, \mnras, 375, 951

\bibitem[2007b]{cameron07b} 
Collier Cameron, A., Wilson, D. M., West, R.G., et al. 2007b, \mnras, 380, 1230

\bibitem[2004]{demarque04} 	
Demarque, P., Woo, Jong-Hak, K., Yong-Cheol, Y., Sukyoung K. 2004, \apjs, 155, 667

\bibitem[2010]{enoch10} 
Enoch, B., Collier Cameron, A., Parley N. R., Hebb L., 2010, \aap, 516, A33

\bibitem[2011]{enoch11} 
Enoch, B., Anderson, D. R., Barros, S., et al. 2011, \apj, 142, 86

\bibitem[2010]{fortney10}
Fortney, J. J., Nettelmann, N. 2010, \ssr, 152, 423

\bibitem[Gillon et al. (2009)]{2009A&A...496..259G} Gillon, M., Smalley, B.,
Hebb, L., et al. 2009, \aap, 496, 259

\bibitem[2008]{gray08} Gray D.F., 2008, The observation and
analysis of stellar photospheres, 3rd Edition (Cambridge University Press),
p.~507.

\bibitem[2009]{hebb09} 
Hebb, L., Collier-Cameron, A., Loeillet, B., et al. 2009, \mnras, 393, 1920

\bibitem[2008]{hebrard08} 
H\'ebrard, G., Bouchy, F., Pont, F., et al. 2008, \aap, 481, 52

\bibitem[2011]{hebrard11} 
H\'ebrard, G., Ehrenreich, D., Bouchy, F., et al. 2011, \aap, 527, L11

\bibitem[1981]{hut81} 
Hut, P. 1981, \aap, 99, 126

\bibitem[Magain (1984)]{1984A&A...134..189M} Magain P. 1984, \aap, 134, 189

\bibitem[2002]{mandel02}
Mandel, K., Agol, E. 2002, \apj, 580, L171

\bibitem[2010]{marigo10} 	
Marigo, P., Girardi, L., Bressan, A., Groenewegen, M. A. T., Silva, L., 
Granato, G. L. 2010, \aap, 482, 883

\bibitem[2008]{matsumura08} 
Matsumura S., Takeda G. \& Rasio F. 2008, \apj, 686, L29

\bibitem[2011]{maxted11} 
Maxted, P. F. L, Anderson, D. R., Collier Cameron, A., et al. 2011,  \pasp, 123, 547

\bibitem[2005]{otha05} 
Ohta, Y., Taruya, A, Suto, Y. 2005, \apj, 622, 1118

\bibitem[2002]{pepe02} 
Pepe, F., Mayor, M., Galland, F., et al. 2002, \aap, 388, 632

\bibitem[2004]{pietrinferni04} 	
Pietrinferni, A., Cassisi, S., Salaris, M., Castelli, F. 2004, \apj, 612, 168

\bibitem[2006]{pollacco06}
Pollacco, D. L., Skillen, I., Collier Cameron, A., et al. 2006, \pasp, 118, 140

\bibitem[2008]{pollacco08} 
Pollacco, D., Skillen, I., Collier Cameron, A., et al. 2008, \mnras, 385, 1576

\bibitem[2000]{queloz00}
Queloz, D., Mayor, M., Weber, L., et al. 2000, \aap, 354, 99

\bibitem[2005]{sato05} 
Sato, B., Fischer, D. A., Henry, G. W., et al. 2005, \apj, 633, 465

\bibitem[Sestito \& Randich 2005]{2005A&A...442..615S} Sestito P., Randich S.
2005, \aap, 442, 615

\bibitem[2009]{skillen09} 	
Skillen, I., Pollacco, D., Collier Cameron, A., et al.  2009, \aap, 502, 391.

\bibitem[2012]{southworth12}
Southworth, J., Hinse, T. C., Dominik, M., et al. 2012, \mnras, 426, 1338

\bibitem[2007]{sozzetti07} 	
Sozzetti, A., Torres, G., Charbonneau, D., et al. 2007, \apj, 664, 1190

\bibitem[2010]{torres10}
Torres G., Andersen J., Gim\'enez A. 2010, \aapr, 18, 67

\bibitem[Torres et al. (2010)]{2010A&ARv..18...67T} Torres G., Andersen J. and
Gim\'{e}nez A. 2010, \aapr, 18, 67

\bibitem[2006]{vandenberg06}
VandenBerg, D. A., Bergbusch, P. A., Dowler, P. D. 2006, \apjs, 162, 375

\bibitem[2009]{west09} 	
West, R. G., Collier Cameron, A., Hebb, L., et al. 2009, \aap, 502, 395

\bibitem[2010]{winn10} 	
Winn, J. N., Fabrycky, D., Albrecht, S., Johnson, J. A. 2010, \apj, 718, L145

\end{thebibliography}
\end{document}